\documentclass[twocolumn,prl]{revtex4}

\usepackage{graphicx}
\usepackage{amssymb}
\usepackage{amsmath}
\usepackage{simplemargins}

\setleftmargin{1cm}
\setrightmargin{1cm}
\settopmargin{1cm}
\setbottommargin{1cm}

\usepackage{color} 

\graphicspath{{NFigs/}{Nov/}}

\newcommand{\iain}[1]{#1}

\makeatletter
\renewcommand{\@biblabel}[1]{\quad#1.}
\makeatother

\date{}

\begin{document}

\begin{flushleft}
{\Large
\textbf{Mitochondrial Variability as a Source of Extrinsic Cellular Noise}
}

Iain G. Johnston$^{1,2}$,
Bernadett Gaal$^{1,3}$,
Ricardo Pires das Neves$^{4,5}$,
Tariq Enver$^3$,
Francisco J. Iborra$^{6, \ast}$,
Nick S. Jones$^{1,2,7, \dagger}$
\\
\bf{1} Department of Physics and CABDyN Complexity Centre, Clarendon Laboratory, Parks Road, Oxford, United Kingdom\\
\bf{2} Oxford Centre for Integrative Systems Biology, Department of Biochemistry, South Parks Road, Oxford, United Kingdom\\
\bf{3} UCL Cancer Institute, University College London, Gower Street, London, United Kingdom \\
\bf{4} Center for Neuroscience and Cell Biology, University of Coimbra, 3004-517 Coimbra, Portugal \\
\bf{5} Biomaterials and Stem Cell-based Therapeutics Group and Biocant -- Center of Innovation and Biotechnology, 3060-197 Cantanhede, Portugal \\
\bf{6} Department of Molecular and Cellular Biology, Centro Nacional de Biotecnolog\'{i}a, Consejo Superior de Investigaciones Cient\'{i}ficas, Madrid, Spain \\
\bf{7} Department of Mathematics, Imperial College London, Huxley Building, 180 Queen's Gate, London, SW7 2AZ, United Kingdom \\
$\ast$ Email: fjiborra@cnb.csic.es \\
$\dagger$ Email: nick.jones@imperial.ac.uk
\end{flushleft}

\section*{Abstract}
We present a study investigating the role of mitochondrial variability in generating noise in eukaryotic cells. Noise in cellular physiology plays an important role in many fundamental cellular processes, including transcription, translation, stem cell differentiation and response to medication, but the specific random influences that affect these processes have yet to be clearly elucidated. Here we present a mechanism by which variability in mitochondrial volume and functionality, along with cell cycle dynamics, is linked to variability in transcription rate and hence has a profound effect on downstream cellular processes. Our model mechanism is supported by an appreciable volume of recent experimental evidence, and we present the results of several new experiments with which our model is also consistent. We find that noise due to mitochondrial variability can sometimes dominate over other extrinsic noise sources (such as cell cycle asynchronicity) and can significantly affect large-scale observable properties such as cell cycle length and gene expression levels. We also explore two recent regulatory network-based models for stem cell differentiation, and find that extrinsic noise in transcription rate causes appreciable variability in the behaviour of these model systems. These results suggest that mitochondrial and transcriptional variability may be an important mechanism influencing a large variety of cellular processes and properties.

\section*{Author Summary}
Cellular variability has been found to play a major role in diverse and important phenomena, including stem cell differentiation and drug resistance, but the sources of this variability have yet to be satisfactorily explained. We propose a mechanism, supported by a substantial number of recent and new experiments, by which cell-to-cell differences in both the number and functionality of mitochondria -- the organelles responsible for energy production in eukaryotes -- leads to variability in transcription rate between cells and may hence be a significant source of cellular noise in many downstream processes. We illustrate the downstream effect of mitochondrial variability through simulated studies of protein expression and stem cell differentiation, and suggest possible experimental approaches to further elucidate this mechanism.

\section*{Introduction}
Stochastic influences significantly affect a multitude of processes in cellular biology \cite{mcadams1997stochastic, altschuler2010cellular, elowitz2002stochastic, kaern2005stochasticity, raj2008nature}. Understanding the sources of this randomness within and between cells is a central current challenge in quantitative biology.  Noise has been found to affect processes including stem cell fate decisions \cite{chang2008transcriptome}, bet-hedging in bacterial phenotypes \cite{fraser2009chance, kussell2005bacterial}, cancer development \cite{brock2009non}, and responses to apoptosis-inducing factors \cite{bastiaens2009systems, spencer2009non}. In this paper, we consider how mitochondria may constitute a significant source of this cellular noise.

Noise in cellular processes may result from sources intrinsic to the gene in question (those responsible for differences in the expression levels of genes under identical regulation in the same cell) or extrinsic sources (those responsible for cell-to-cell variation in genes under identical regulation in a population). Both intrinsic and extrinsic noise sources contribute to the overall noise observed in, for example, transcription rates and protein expression levels \cite{swain2002intrinsic}. The interplay between intrinsic and extrinsic noise can be characterised with elegant experimental techniques such as dual reporter measurements \cite{elowitz2002stochastic}, in which the expression levels of two proteins are measured within cells and within a population, but subtleties exist in disambiguating intrinsic and extrinsic contributions to noise levels \cite{hilfinger2011separating}. Some studies have found the contribution of extrinsic factors to overall noise levels to be stronger in eukaryotes \cite{raser2004control,newman2006single} than prokaryotes \cite{elowitz2002stochastic}, although others debate this interpretation \cite{raj2006stochastic}. To investigate these influences, several mathematical models for the emergence of intrinsic and extrinsic cellular noise have been introduced and explored \cite{paulsson2005models, paulsson2004summing, volfson2005origins, bruggeman2009noise, rausenberger2008quantifying, blake2003noise, dobrzynski2009elongation, swain2002intrinsic, thattai2001intrinsic}. In addition, recent studies have investigated, both experimentally and theoretically, the architecture of extrinsic noise and its causal factors \cite{raser2004control, sigal2006variability, sigal2006dynamic, volfson2005origins, bareven2006noise, raj2006stochastic, newman2006single}, though substantial uncertainty surrounds the importance of individual contributions (such as variability in cell cycle stage and cellular volume) to extrinsic noise \cite{kaufmann2007stochastic}.  

Huh and Paulsson recently argued that uneven segregration of cellular constituents at mitosis can contribute significantly to cell-to-cell differences in levels of cellular components and proteins in a population, focusing on stochasticity in protein inheritance between sister cells \cite{huh2010non, huh2011random}. We focus on a specific instance of this phenomenon: cell-to-cell variability in the mitochondrial content of cells. An experimental study performed by das Neves \emph{et al.} identified uneven partitioning of mitochondria at mitosis as being a possibly significant source of extrinsic noise in eukaryotes \cite{neves2010connecting}, supporting recent theoretical ideas \cite{huh2011random}. Mitochondria have been found to display remarkably complex behaviour interwoven with cellular processes \cite{mcbride2006mitochondria, chan2006mitochondria, twig2008mitochondrial} and to display significant heterogeneity within cells \cite{neves2010connecting, collins2002mitochondria, buckman2001spontaneous, oreilly2003quantitative}. Mitochondrial influences on processes including stem cell differentation \cite{schieke2008mitochondrial} and cell cycle progression \cite{mitra2009hyperfused, mandal2005mitochondrial, owusu2008distinct} have recently been observed.

\iain{das Neves \emph{et al.} \cite{neves2010connecting} observe a wide spread of mitochondrial masses in a population of cells, illustrating extrinsic variability in organelle distribution. Mitochondrial functionality has also been observed to vary between cells \cite{collins2002mitochondria, twig2008mitochondrial, kuznetsov2006mitochondrial, cossarizza1996functional, mouli2009frequency}. das Neves \emph{et al.} also observed a link between mitochondrial mass and membrane potential and cellular ATP levels, and found transcription rate to be a function of ATP concentration. In addition, the modulation of mitochondrial functionality, through anti- and pro-oxidant treatments, was found to alter cell-to-cell variability in transcription rates, with anti-oxidants significantly reducing variability and pro-oxidants increasing variability.  These results suggest that cell-to-cell heterogeneity in mitochondrial mass and functionality may propagate into extrinsic noise in transcription rate, and thenceforth processes further downstream, but the quantitative links behind these processes remain unclear. We introduce a simple approach, consistent with a range of experimental observations, that quantitatively connects all these features and predicts the downstream physiological influence of mitochondrial variability.}

\iain{Shahrezaei \emph{et al.} \cite{shahrezaei2008colored} have recently shown that extrinsic noise can influence levels of intrinsic noise, as cell-to-cell variability in the rates of processes such as transcription and translation affect the intrinsic dynamics of gene expression. \iain{In addition, they provided an extension to standard stochastic simulation techniques to allow this variability in the production rates of chemical species to be accurately simulated -- a problem that has been approached using different techniques in previous studies \cite{jansen1995monte, haseltine2002approximate}. However, this theoretical study did not attempt to characterise the physiological causes of this extrinsic noise -- an important consideration in assessing the ubiquity and consequences of cellular noise. Our proposal that cell-to-cell mitochondrial variability provides a significant source of extrinsic noise in transcription addresses these causes, and we show that extrinsic noise resulting from mitochondrial variability could significantly influence intrinsic noise in gene expression.}} 

This paper will proceed as follows. We first introduce one of the simplest possible mathematical models for variation in mitochondrial mass and functionality during and between cell cycles, and show that it is consistent with a wide range of experimental data, both from the literature and newly reported here, and allows analytical treatment. Our model includes stochastic segregation of mitochondria at mitosis and functional differences in mitochondria between cells, and contains a simple dynamic description of the time evolution of cellular volume and mitochondrial mass through the cell cycle. \iain{To our knowledge it is the first model of its kind which links mitochondrial mass and function to the cell cycle and gene expression. We relate mitochondrial properties} to the production of ATP in the cell, which in turn affects transcription rates: hence, variability in mitochondrial properties causes downstream variability in transcription. Next, we incorporate the behaviour produced by our model into a common framework for cellular noise, and \iain{show that extrinsic noise due to variation in $[ATP]$ can have a profound effect on gene expression levels, dominating over intrinsic noise}. We then demonstrate the cell physiological implications of energy variability by showing how mitochondrial variability may affect stem cell differentation. Finally, we discuss how our model relates to recent work characterising sources of extrinsic noise, and suggest experiments to allow more refined models.

\section*{Model}
While the heterogeneity of mitochondria has been observed experimentally and connected to variability in processes like transcription \cite{neves2010connecting} and stem cell differentiation \cite{schieke2008mitochondrial}, the mechanisms by which mitochondrial variability influences other cellular processes has not been elucidated clearly. Here, we describe a simple model which formalises these links, and note that it reproduces recent experimental results concerning mitochondrial heterogeneity (and variability in connected cellular features) \cite{neves2010connecting}. The simplicity of our model means that analytic expressions can be derived for many quantities of interest, facilitating a more complete and intuitive understanding of the modelled biological connections. We will then use this model to investigate more specific questions regarding the links between mitochondrial variability and transcription rate and stem cell differentiation. 

\begin{figure}
\includegraphics[width=9cm]{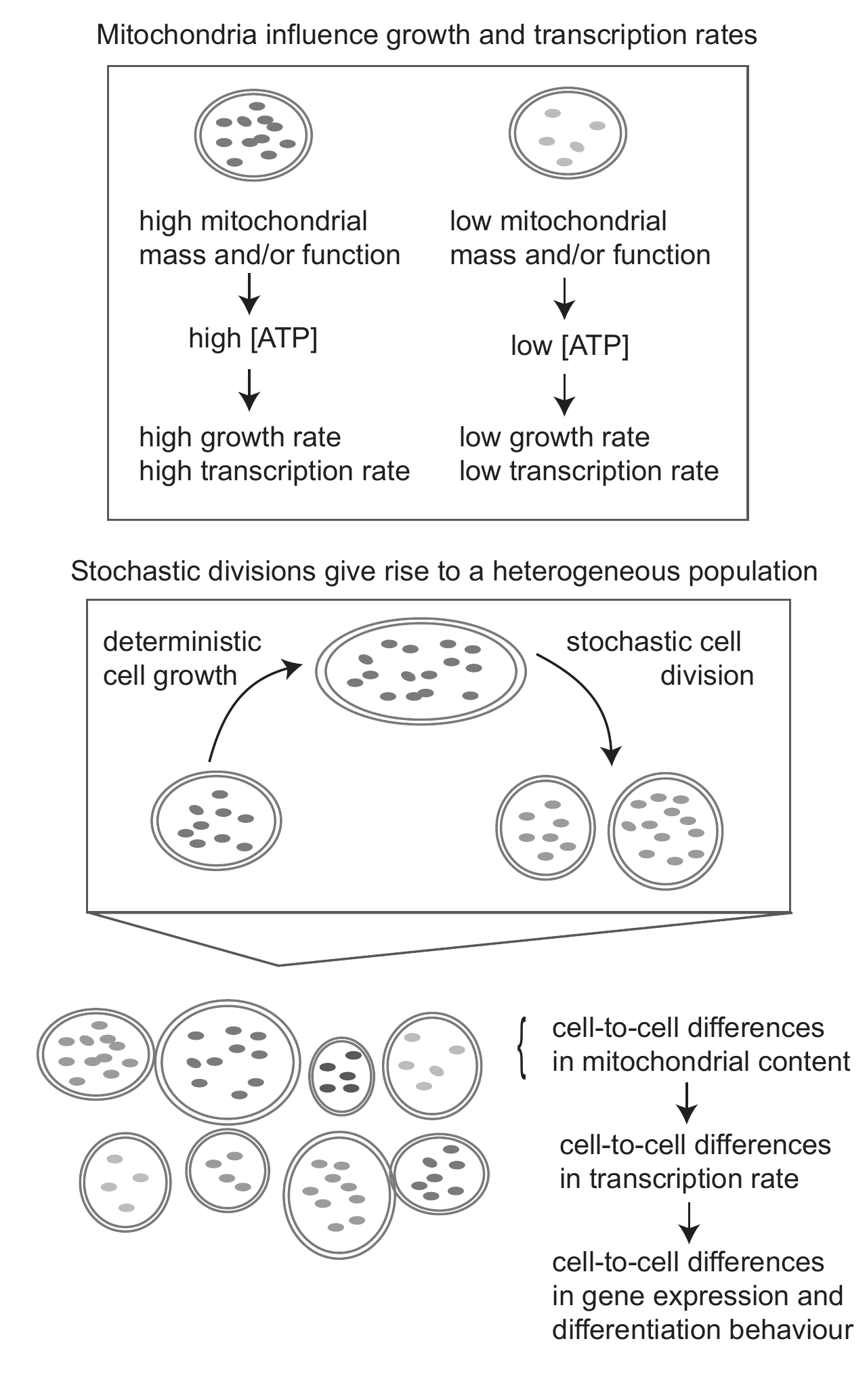}
\caption{\textbf{An illustration of the model we employ for mitochondrial variability.} This illustration qualitatively shows the key components of our model. Cell growth progresses deterministically according to the variables that characterise a cell: volume, mitochondrial mass (illustrated here by copy number) and functionality (illustrated here by shading). At mitosis, stochastic partitioning occurs and daughter cells inherit a random volume, mitochondrial mass and functionality level from a parent cell. This stochastic inheritance leads to a heterogeneous population. Cells with high mitochondrial density and functionality have higher ATP levels, are able to grow faster, and have higher transcription rates than cells with lower mitochondrial mass and functionality. The variances associated with stochastic partitioning, the dependence of ATP concentration on cellular properties, and the dependence of growth and transcription rates on ATP are all parameters of the model.}
\label{summaryfig}
\end{figure}

The central concept behind our model is illustrated in Fig. \ref{summaryfig}. Individual cells are characterised by three key variables: the volume of the cell ($v$); the amount of mitochondrial mass in the cell ($n$); and the degree of mitochondrial functionality ($f$). This last quantity, $f$, represents a coarse-grained measure of the efficiency of mitochondria within a cell -- a factor which may be affected, for example, by the levels of reactive oxygen species (ROS), mitochondrial membrane potential, variability in mitochondrial protein complex abundance, and genetic differences between mitochondria \cite{lane2006mitochondrial}. ATP concentration in the cell is modelled as a function of these three quantities, and transcription rate is modelled as a function of ATP concentration \cite{neves2010connecting}. Variability in cell volume, mitochondrial mass and mitochondrial functionality arises due to stochastic inheritance of these quantities at cell divisions. This variability causes cell-to-cell differences in ATP levels, and hence transcription rate, in a population of cells.

\subsection*{Stochastic Partitioning at Mitosis}
In our model, cells grow deterministically (see Cellular Dynamics), undergoing mitosis when their volume reaches a cutoff $v^*$. When this occurs, the cell divides in two, with mitochondrial mass $n$ split stochastically between daughter cells, with each unit of mass being assigned to each cell with equal probability, and cell volume also segregated randomly (see Methods).  In our model, the partitioning of $n$ and $v$ at mitosis is uncorrelated. We use this lack of correlation both for simplicity and due to experimental data (see Results) illustrating that cell cycle length correlates well with inherited mitochondrial mass and poorly with inherited cell volume, indirectly suggesting a lack of correlation between $n$ and $v$. This model was chosen as the most straightforward representation of stochastic division of discrete elements, and is likely to represent a realistic scenario if there is no explicit biological control mechanism that modulates the distribution of inherited mitochondria.


Our mitochondrial mass measure $n$ physically represents total mitochondrial volume. However, it will be of use when considering the segregation of mitochondria at mitosis to consider the cell as populated by a number of discrete `virtual' mitochondria.  We denote these entities as `virtual' mitochondria due to the difficulty of regarding mitochondria as individuals given the processes of fission and fusion \cite{scott2010mitochondrial}. The system as chosen is scaled so as to regard $n$ as mitochondrial copy number, so that, if $n$ is measured in $\mu m^3$, each `virtual' mitochondrion possesses a default volume of $1\,\mu m^3$ (see Methods). These virtual mitochondria are the discrete elements that, in our model, are binomially partitioned at mitosis. We use the binomial picture both for simplicity and due to its agreement with recent data on mitochondrial partitioning \cite{neves2010connecting}, but note that a range of mitochondrial partitioning regimes have been observed in the literature \cite{huh2010non, wilson1916distribution, wilson1931distribution}, and explore (in the Results section) the effects of wider or narrower distributions associated with mitochondrial partitioning.

We consider the variable $f$ to be the degree of functionality of a cell's mitochondria. The inclusion of such a term is necessitated by several experimental observations. das Neves \emph{et al.} show that a measure of mitochondrial functionality (membrane potential) is slowly-varying with time in a given cell, although there is a wide distribution of functionality within a population \cite{neves2010connecting}. It was also found that sister cells have similar transcriptional noise levels compared to the bulk population: if stochasticity were to arise from mitochondrial mass partitioning alone, we would expect sister cells, post-mitosis, to exhibit the greatest possible variation, as subsequent cell growth may be expected to dampen such variability \cite{huh2011random}. Another experimental observation is that populations of cells treated with antioxidants, which improve mitochondrial functionality, showed a significant drop in noise levels. These results suggest that an extra source of noise, functional variability between cells, may be responsible for increasing noise levels. 

In the absence of a more refined view of functionality, we assume that all changes in functionality occur at division and that $f$ stays constant through the cell cycle. $f$ changes in a stochastic but mean-reverting fashion at division, and both daughters receive the same $f$ value (see Methods for more detail). In this simple model, the variation that a cell experiences due to slow changes in mitochondrial functionality through the cell cycle is absorbed into stochastic changes at cell division. We choose this modelling protocol due to the absence of detailed data on the behaviour of mitochondrial functionality on timescales longer than a cell cycle, and suggest that parameterising this simple system to match the experimentally observed distribution of mitochondrial functionality will give a reasonable estimate of the population variability in this quantity. In `Other Models' in Supplementary Information, we discuss another picture in which we allow $f$ to vary continuously through the cell cycle, and show that similar results emerge when this alternative model is used.

In this study, we will consider the oxidative state of a cell as a key mediator of its functionality $f$. Recent experimental data has shown that treating cells with pro- or anti-oxidants strongly affects the statistics of transcription rate variability in a population \cite{neves2010connecting}. Within our model, the effects of such chemical treatments on the oxidative state of cells can straightforwardly be captured by varying the parameters associated with functional inheritance (see Methods).


\subsection*{$[ATP]$ and Transcription Rate}

We are interested in the time evolution of $[ATP]$ as a potential stochastic influence on downstream processes. Ref. \cite{neves2010connecting} found ATP levels in the cell to be proportional to mitochondrial mass ($n$) and membrane potential (a factor that may be absorbed into our measure of `mitochondrial function' $f$), motivating our choice of expression for ATP concentration:

\begin{equation}
[ATP] = \frac{\gamma n f}{v}.
\label{firsteqn}
\end{equation}

In this expression, $\gamma$ is a constant of proportionality linking the quantities within our model to a biological ATP concentration, and the meaning of the variable $f$ now becomes apparent as a scalar multiple relating mitochondrial density to $[ATP]$. We note that other choices for the form of  $[ATP]$, including ODEs, are possible, and explore some alternatives in `Other Models' (Supplementary Information). das Neves \emph{et al.} also show a link between the total transcription rate $\lambda$ in a cell (measured through bromo-uridine incorporation across the whole nucleus) and $[ATP]$, a sigmoidal curve, which we approximate (see `Parameterisation of $\lambda$' in Supplementary Information) with 

\begin{equation}
\lambda = s_1 + s_2 \tan^{-1} \left( s_3 [ATP] + s_4 \right).
\end{equation}

das Neves \emph{et al.} record a change in the structure of this sigmoid curve in experiments where cellular chromatin is artifically decondensed. In these situations, the sigmoidal response of $\lambda$ to $[ATP]$ becomes a hyperbolic curve, with a sharp, continuous increase of $\lambda$ with $[ATP]$ at low $[ATP]$. This change may reflect the necessity of remodelling chromatin -- a process that requires ATP -- for the transcription process. Chromatin remodelling has been noted by several studies \cite{newman2006single, blake2003noise, raj2006stochastic} to play an important role in mRNA synthesis noise and hence downstream noise in gene expression. Rather than attempting to model this influence explicitly, we use the experimentally-determined form for $\lambda([ATP])$ to capture the overall dependence of transcription rate (including chromatin effects) on $[ATP]$.

To summarise, in our model, transcription rate depends sigmoidally on ATP concentration -- a relationship elucidated and quantified in recent experiments \cite{neves2010connecting}. ATP concentration in turn depends linearly on the mitochondrial mass and functionality level of a cell and also on the cell volume. Cells with many, highly functional mitochondria will have higher levels of ATP and hence higher transcription rates than those with smaller, less functional mitochondrial populations.

\subsection*{Cellular Dynamics}
Our model for cell cycle dynamics consists of equations governing the time evolution of the key quantities volume, mitochondrial mass, and mitochondrial functionality. In the light of a recent study \cite{tzur2009cell}, and as cell cycle models often assume the exponential growth picture, we expect an exponential form for cell volume growth: $\dot{v} \propto v F(v, n, f)$. Here, $F(v, n, f)$ is a function expressing the dependence of volume growth rate on other parameters. 

We suggest that ATP concentration ($[ATP]$) plays a key role in powering growth of the cell, so cells with higher ATP levels have higher growth rates associated with cell volume and mitochondrial mass. This link postulates that biosynthesis rates are generally, like transcription, a function of ATP concentration. We note that although ATP concentration has been suggested \cite{neves2010connecting} as a possible mechanism linking mitochondria and transcription rate, and some evidence supports this link, it may be the case that a different factor provides the causal mechanism, and ATP concentration is correlated with this underlying factor. For example, ROS, which adversely affect many cellular processes, may be an alternative to ATP, or a combination of ATP and ROS levels may act to determine transcription rate.


Numerous historical studies, both in HeLa cells \cite{posakony1977mitochondrial} and other tissue types \cite{veltri1990distinct, herbener1976morphometric, mathieu1981design, suarez1991mitochondrial, hoppeler1984scaling, robin1988mitochondrial} have found that the density $\rho$ of mitochondrial mass (also called mitochondrial volume density) within cells of a given tissue type is consistent between generations and within populations. This consistency suggests that the time evolution of mitochondrial mass should be (a) coupled with the time evolution of volume and (b) of a form that allows damping of the inherent stochasticity at mitosis. In addition to these features, it is presumably reasonable to assume that mitochondrial growth is dependent on available $[ATP]$ (due to the required protein synthesis). We suggest a model that captures these required dependencies and incorporates mean-reversion, given by the dynamic equations: 

\begin{eqnarray}
\dot{v} & =  & \alpha f \rho v \\
\dot{n} & = & \beta f \rho v, \label{lasteqn}
\end{eqnarray}

where $\rho = n/v$. 

We note that this simple model does not distinguish between volume growth rates at different times in the cell cycle, but yields a smooth exponential growth in cell size throughout the cell cycle. We work in this picture for simplicity and generality, but note that a more sophisticated model would include a more detailed description of cell growth as another potential source of variability between cells.

The model's dynamics result (see Methods) in a convergence in mitochondrial density with time to a value $\beta/\alpha$.

\subsection{Model Parameterisation}
\begin{figure}
\includegraphics[width=9cm]{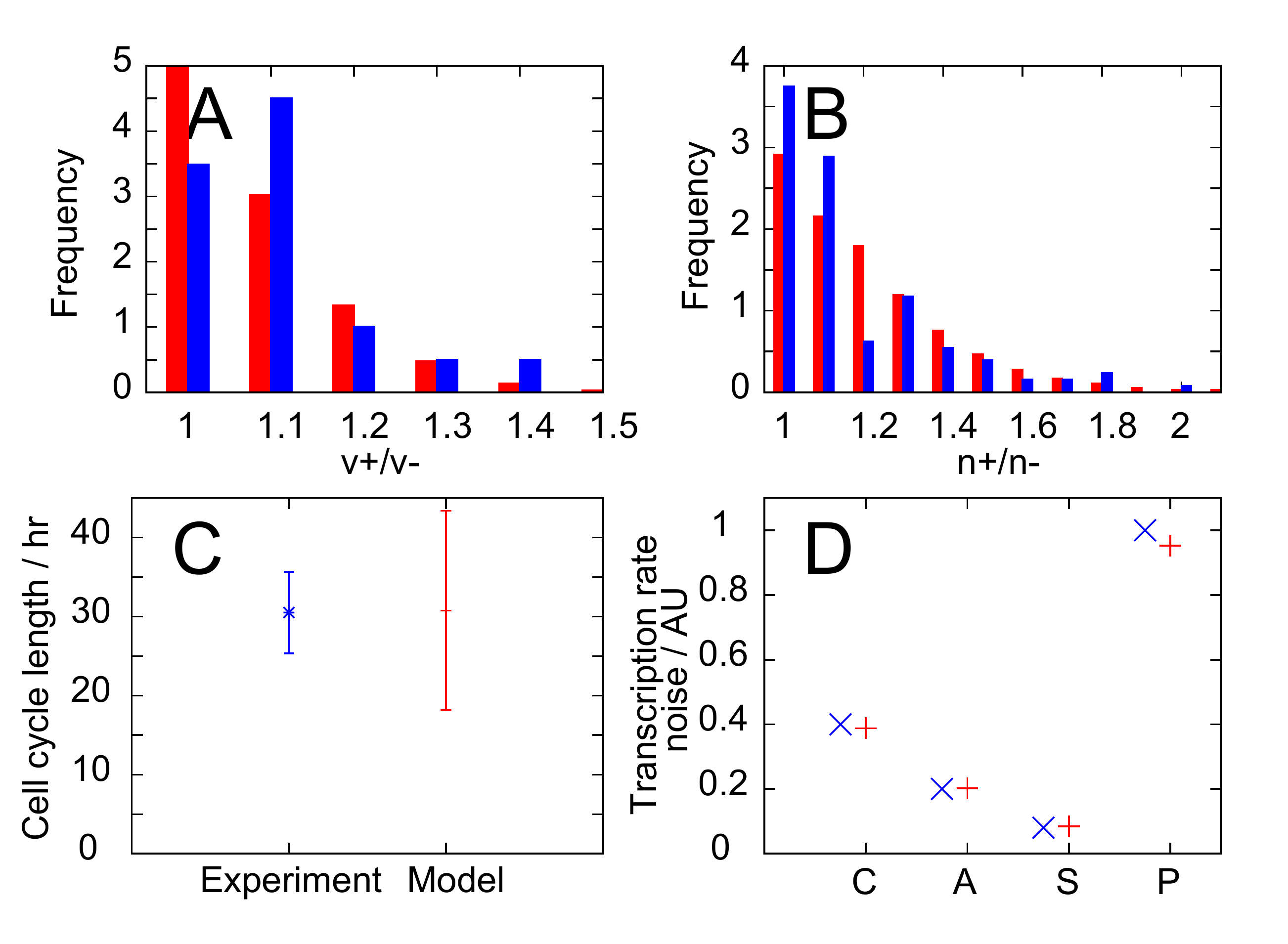}
\caption{\textbf{The set of data used to parameterise our model.} Experimental data shown in blue, fitted simulated data shown in red. A. Ratio of larger cell volume to smaller cell volume between sisters at birth.  B. Ratio of larger mitochondrial mass to smaller mitochondrial mass between sisters at birth. C. Mean and standard deviation of the cell cycle length in a population of cells. D. Noise levels in transcription rate in (C)ontrol, (A)ntioxidant-treated and (P)ro-oxidant-treated populations, and between (S)ister cells.  Two other experimental values, not pictured, that were used to parameterise our model are a maximum cell volume of $2\,500\,\mu$m$^3$ (for consistency with Ref. \cite{tzur2009cell}) and a mean ATP concentration of $900\,\mu$M (from Ref. \cite{wang1997turnover}).}
\label{wufig1}
\end{figure}

Values for the parameters in our model were chosen (see Methods) to match a subset of experimental data, illustrated in Fig. \ref{wufig1}. 

\section*{Results}
In this section, we first compare recent experimental data to the predictions of our model and demonstrate that a good agreement exists across a wide range of experiments. We next report new experimental results of relevance to the study of mitochondrial variability and show that these too largely agree with the predictions from our simple model. This set of successful comparisons suggests that our model is capable of producing quantitatively sound estimates of the levels of noise associated with mitochondrial sources of variability. Motivated by these results, we next show how our model allows a quantitative link to be formed between mitochondrial variability and variability in transcription rate in cells. We explore this link by investigating the predictions that our model makes concerning noise in models of gene expression levels, and in models of stem cell differentiation pathways. We find that the mitochondrial sources of variability from our model provide a substantial contribution to noise levels in mRNA and protein levels within the cell, and can influence stem cell differentiation in a manner that depends upon the symmetry of the regulatory interactions that drive differentiation.

\subsection*{\iain{Our simple model is sufficient to approximate a large set of experimental data}}
Here we list a set of comparisons between predictions from our model and experimental studies. Unless stated otherwise, we will use experimental data from the study of das Neves \emph{et al.} \cite{neves2010connecting}, using the protocol `NX' to refer to data in Fig. X of that study.

\begin{figure*}
\includegraphics[width=16cm]{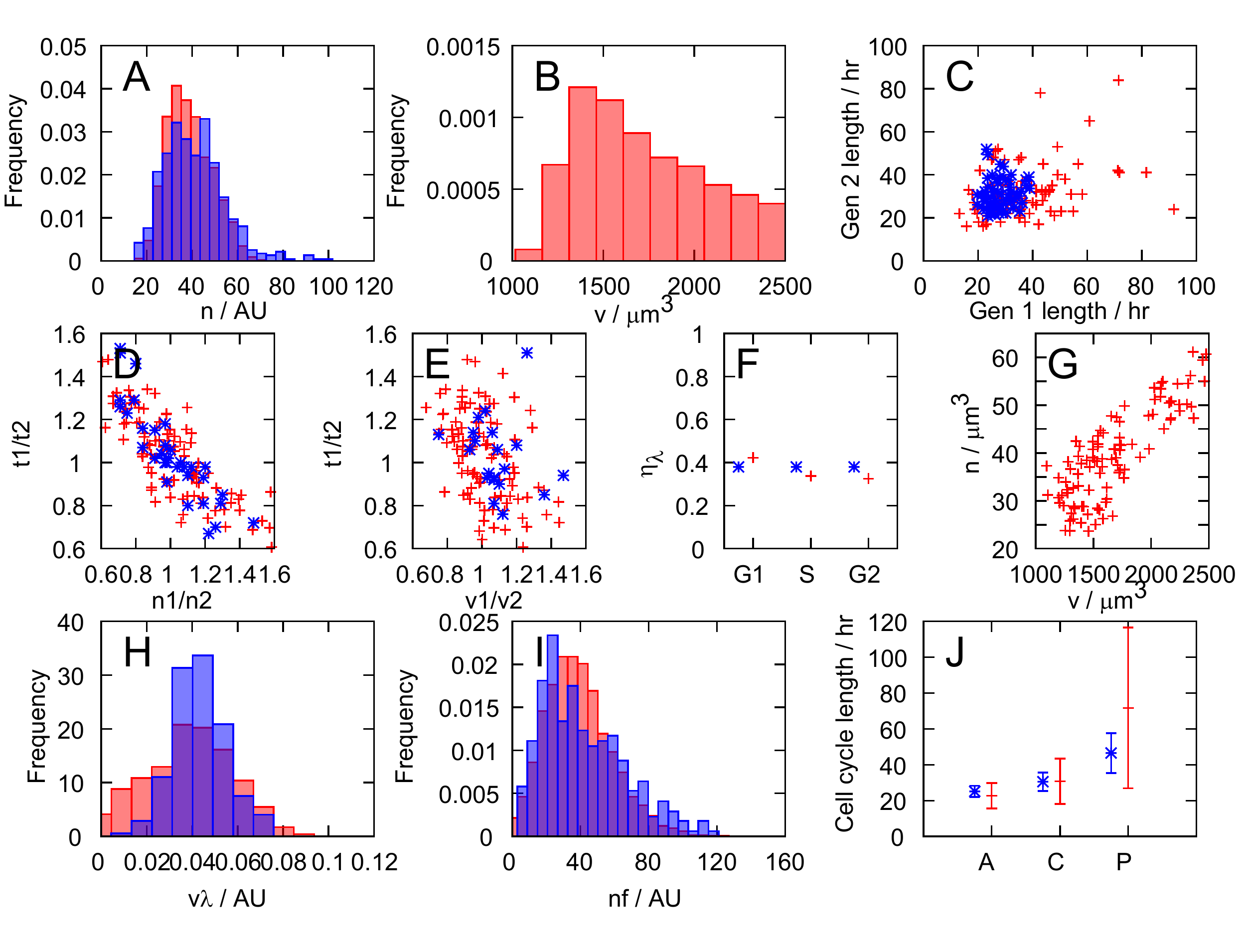}
\caption{\textbf{Our simple model is consistent with experimental probes of mitochondrial and cellular variability.} Comparison between our model (red) and experimental data (blue), following discussion in the Main Text. \textbf{Experimental data from das Neves \emph{et al.} \cite{neves2010connecting}.} A. Distribution of mitochondrial mass $n$ in an unsynchronised population of cells. B. Distribution of cell volume $v$ in an unsynchronised population of cells. C. Comparison of the lengths of cell cycles between generations: Gen 1 is the parent cell, Gen 2 the daughter. Cell cycle lengths are only weakly correlated. D. Relationship between the ratio of mitochondrial masses at birth against ratio of cell cycle lengths for sister pairs. E. Relationship between the ratio of cellular volumes at birth and the ratio of cell cycle lengths for sister pairs, showing a weaker correlation than D. F. Transcription rate noise $\eta_\lambda$ in subsets of the population in $G_1$, $S$, and $G_2$ phases (see Main Text). G. Mitochondrial mass $n$ and cell volume $v$ are strongly correlated in our model. Some experimental evidence is contradictory (see Main Text).  H. Distribution of transcription rate per unit volume $\lambda/v$. \textbf{New experimental data (see Methods).} I. Distribution of total mitochondrial functionality ($nf$ in our model, CMXRos readings from experiments). J. Mean and standard deviation of cell cycle lengths in (A)nti-oxidant-treated, (C)ontrol, and (P)ro-oxidant-treated populations. Experimental histograms, originally presented in arbitrary units, have been scaled to match the mean value of the simulated data.}
\label{wufig2}
\end{figure*}

\emph{Distributions of mitochondrial mass and cell volume.} Our model gives a peaked distribution skewed towards low $n$ values for mitochondrial mass in the bulk population (Fig. \ref{wufig2}A), which is similar in form to the experimental distribution (N4b). The distribution of cellular volumes in a bulk population (Fig. \ref{wufig2}B) is found to display the quadratic decay expected from a theoretical treatment of cells growing exponentially \cite{volfson2005origins}. 

\emph{Weak correlation between the lengths of successive cell cycles in a population}. Fig. \ref{wufig2}C shows the weak relationship between the cell cycle length of a parent and a daughter cell, which qualitatively matches experimental findings (from N4h).

\emph{Mitochondrial mass at birth is a better predictor of cell cycle length than cell volume at birth.} Figs. \ref{wufig2}D and \ref{wufig2}E illustrate the correlations between cell cycle length and a cell's birth values of $n$ and $v$ respectively. The correlation between birth mitochondrial mass and cell cycle length was strong ($R^2 = 0.69$, compared to the experimental value of 0.78) compared to the correlation between birth cell volume and cell cycle length ($R^2 = 0.22$, experimental value 0.22). The same correlation behaviour is observed in experiments (from N4e and N4f) which are shown for comparison. 

\emph{Transcription rate noise with cell cycle stage. } We modelled progression through the cell cycle stages by assigning stages according to the volume $v$ of a cell. We assign cells with $0.5 v^* \leq v < 0.7 v^*$ to $G_1$, $0.7 v^* \leq v < 0.95 v^*$ to $S$, and $0.95 v^* \leq v < v^*$ to $G_2$ stages, to approximate the proportion of total cell cycle length that HeLa cells are observed to spend in each stage \cite{kumei1989reduction}. Transcription rate noise was found to stay relatively constant (around 0.4) when population subsets at different positions in the cell cycle were measured (see Fig. \ref{wufig2}F), as observed in experiments (NS1).

\emph{Correlation between mitochondrial mass and cell volume.} Our model predicts a strong correlation between cell volume $v$ and mitochondrial mass $n$ (Fig. \ref{wufig2}G). This result contrasts with the weak correlation observed, using forward scatter in flow cytometry to measure volume, by das Neves \emph{et al.} (N3a) (we confirmed these experimental results in this study -- data not shown). However, many historic studies have found a much stronger connection between mitochondrial mass and cellular volume. The mitochondrial density $\rho = n/v$, also referred to as mitochondrial volume density, has been found to exhibit low standard deviation (between 0.01 and 0.15 of the mean) in many different mammalian tissue types \cite{veltri1990distinct, herbener1976morphometric, mathieu1981design, suarez1991mitochondrial, hoppeler1984scaling, posakony1977mitochondrial, robin1988mitochondrial} and amounts of mtDNA have been found to display similarly low variability \cite{veltri1990distinct, bogenhagen1974number}. These results contrast with the extremely high variability in mitochondrial volume density observed by das Neves \emph{et al.} (the noise level estimated from the data is around 0.32), but we note that flow cytometry data (while useful for providing approximate orderings of cells by volume) may not be capable of providing the absolute volume measurements which are required to refute the low variability in $\rho$ observed in many other studies. 


\emph{Distribution of transcription rate per unit volume.} Fig. \ref{wufig2}H shows the distribution of transcription rate per unit nuclear volume (in our model, nuclear volume is taken as proportion to cell volume) in the bulk population. This result follows a similar peaked distribution to that found experimentally (N1a). 

\begin{figure}
\includegraphics[width=9cm]{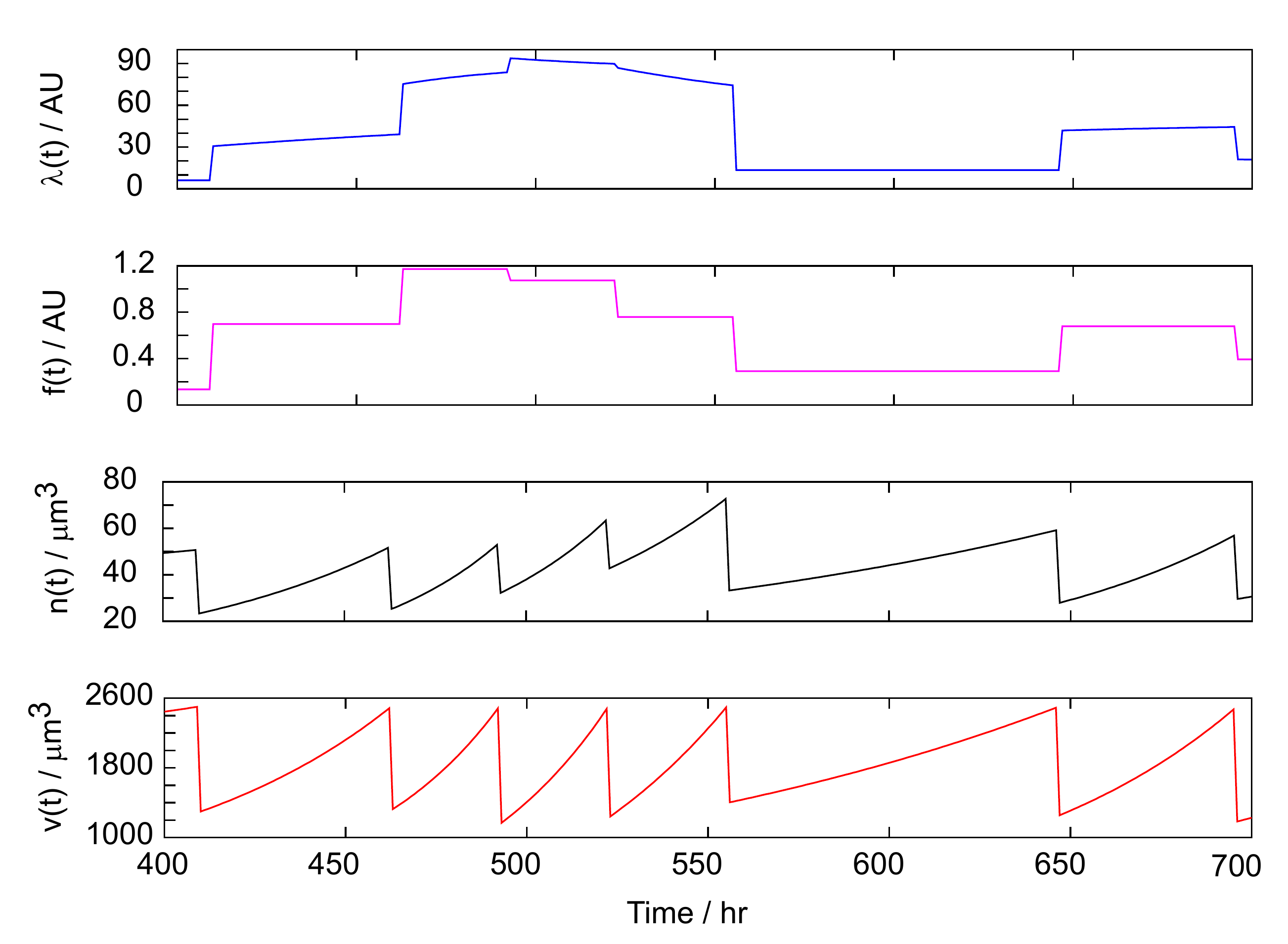}
\caption{\textbf{Illustration of the dynamics of our model.} Example time series of $\lambda$ (transcription rate), $f$ (mitochondrial functionality), $n$ (mitochondrial mass) and $v$ (cell volume), as a cell grows and divides repeatedly in our model.}
\label{wufig3}
\end{figure}

\emph{Others.} We also note some qualitative features of our model: an increase in transcription rate with ATP levels is observed (trivially due to the functional form of $\lambda$), which is also observed experimentally (N3g). We also observe an increase in transcription rate per unit volume with total mitochondrial functionality ($nf$ in our model), found experimentally (N3d). Fig. \ref{wufig3} shows illustrative time series of the dynamic variables involved in simulation of our model.

\subsection*{\iain{New experimental results are also consistent with this model}}
In Fig. \ref{wufig2}, we also present new experimental results pertaining to our model. These new experiments were designed to characterise two additional features of cells in a population: a measure of the total level of mitochondrial function within cells and the modulation of cell cycle lengths by changing the oxidative state of the cell. The total level of mitochondrial function is experimentally measured using the intensity of signal from CMXRos, a dye that stains mitochondria and accumulates according to membrane-potential, integrated over a whole cell (see Methods). This signal reports on the integrated membrane potential across the entire cell, combining measures of mitochondrial mass and functionality. The population distribution of this quantity is of interest in exploring the link between mitochondrial mass and functionality between cells.

The modulation of cell cycle length with cellular oxidative state was investigated by observing the distribution of cell cycle lengths in a control population of cells and in populations of cells after anti-oxidant (dithiothretiol) or pro-oxidant (diamide) treatments (see Methods). Our model incorporates oxidative status by modulating the mean level of mitochondrial functionality, so mitochondria function more readily in an environment with low oxidative stress than one with high oxidative stress. As mitochondrial functionality is tied in our model, through growth rate, to cell cycle length, we would expect cell cycle lengths to decrease upon anti-oxidant treatment and increase upon pro-oxidant treatment.

\emph{Distribution of mitochondrial functionality.} Fig. \ref{wufig2}I shows the distribution of total mitochondrial functionality in a population of cells. In our model, this distribution is just the distribution of the quantity $nf$, and in experiments, we measure the total membrane potential within a cell (see Methods). The predicted and experimentally observed distributions share a skewed form with similar variances. 

\emph{Cell cycle lengths in different oxidative conditions.} In Fig. \ref{wufig2}J we show the mean and standard deviation of cell cycle lengths in a control population, and upon treatment with anti- and pro-oxidants (see Methods). In our simulations, these treatments are modelled by changing the value of $f_c$, affecting the mean functionality of mitochondria (see Table \ref{paramvals}). It is observed that treatment with anti-oxidants reduces cell cycle lengths, and treatment with pro-oxidants increases cell cycle lengths. In our model, this behaviour emerges from the dependence of the rate of volume growth on $[ATP]$, and the increased $[ATP]$ levels resulting from mitochondria with higher functionality.

\emph{Mitochondrial mass and membrane potential.} We also observed a linear correlation between total mitochondrial mass (measured with MitoGreen) and total mitochondrial membrane potential (measured with CMXRos) in experiments performed with both dyes (see Methods and `Mitochondrial Membrane Potential' in Supplementary Information). This linear correlation emerges from our model due to our representation of total mitochondrial functionality as the product of a functional measure $f$ with mitochondrial mass $n$. The observed correlation provides qualitative support for this representation.


\subsection{Summary of comparisons between experimental results and model predictions}

It can be seen that several key experimental results require the inclusion of terms relating to mitochondrial variability for an explanation. In a situation without considerable mitochondrial influence on cellular variability, it may be expected that variability in cell cycle position among a population of unsynchronised cells may be a dominant source of noise. Physical distributions subject to such cell cycle noise would be expected to show a variance over around an approximately twofold range, as this is the maximum difference in size between two unsynchronised cells. However, several results display data that varies over a considerably wider range than a factor of two, indicating that a factor other than cell cycle variability may be responsible. Most straightforwardly, Figs. \ref{wufig2}A and \ref{wufig2}I demonstrate pronounced cell-to-cell variability in the mass and functionality of mitochondrial populations. The distribution of transcription rate in Fig. \ref{wufig2}I similarly shows a wide range of values. 

Figs. \ref{wufig2}D and \ref{wufig2}E demonstrate the observed fact that mitochondrial inheritance at birth is a better predictor of cell cycle length than volume inheritance: an effect that relies on the presence of mitochondrial variability and mitochondrial influence on cellular growth. The variability in cell cycle length observed by modulating the oxidative state of the cell in Fig. \ref{wufig2}J suggests that a source of variability that is sensitive to oxidative effects strongly affects cell cycle lengths. We believe that these results support the hypothesis that mitochondrial variability provides a significant contribution to the variability in distributions of the cellular properties we consider.

The correspondence between experimental data and the simulated behaviour of our model suggests that, although we have chosen simple functional forms in our model, the resulting behaviour is biologically relevant. However, we note here that our model was constructed from a phenomenological philosophy, with the intention of using experimental results to construct a plausible coarse-grained explanation for the influence of mitochondrial variability on extrinsic noise in general and transcription rate in particular. Our goal was to introduce a simplified but consistent mathematical summary of the data and to use this to motivate further experiments. To this end, we suggest a set of experiments in `Potential Experiments for Refinement' (Supplementary Information) that would support or contribute to further development of this model. We also note that many potential refinements could be made to our model and suggest several other functional forms in `Other Models' (Supplementary Information).




\subsection*{\iain{Noise in transcription rate depends on noise in mitochondrial segregation and functionality}}
We are now in a position to explore the dependence of the level of noise in transcription rate on the stochasticity in mitochondrial mass and function, and subsequent stochasticity in $[ATP]$. To investigate the contribution of mitochondrial variability sources to transcription rate noise, we performed simulations of our model while varying $\sigma^2_f$, the variance associated with the inheritance of mitochondrial functionality, and $\sigma^2_n$, the variance associated with inheritance of mitochondrial mass. $\sigma^2_n$ here gives the variance of the distribution by which mitochondrial mass is partitioned, and varying it under the assumption of binomial partitioning corresponds to changing the mitochondrial makeup of the cell: lower $\sigma^2_n$ corresponds to more mitochondrial elements, each with smaller volume, while higher $\sigma^2_n$ corresponds to fewer, larger mitochondrial elements, which are partitioned binomially at mitosis (see Methods). 

\begin{figure}
\includegraphics[width=9cm]{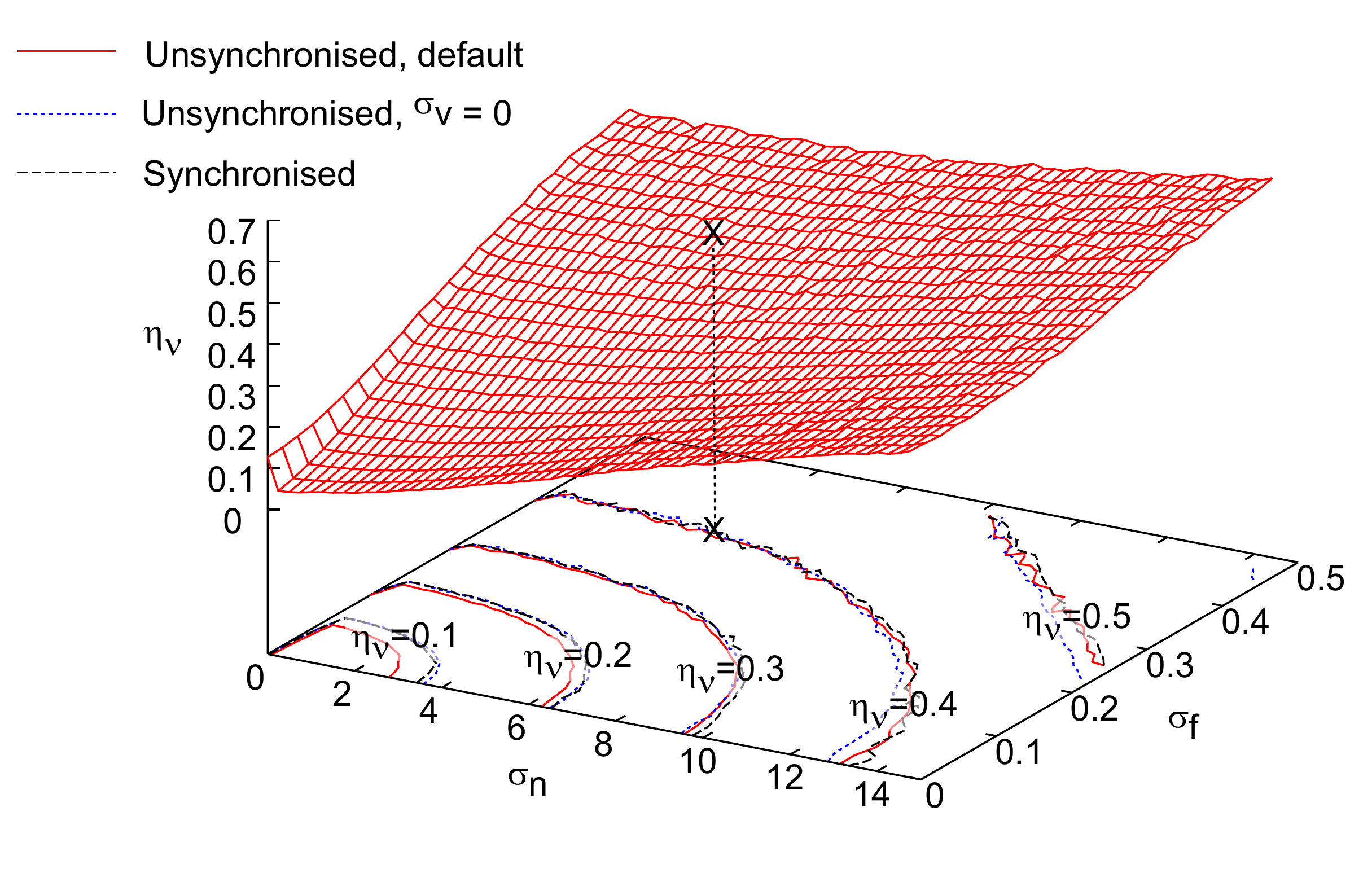}
\caption{\textbf{Variability in mitochondrial mass and functionality can both contribute to noise in transcription rate.} Effects of changing variability in mitochondrial mass inheritance ($\sigma_n$) and functionality ($\sigma_f$) on overall transcription rate noise $\eta_{\lambda}$. This contour plot shows the value of $\eta_{\lambda}$ for a given combination of $\sigma_n, \sigma_f$. More stochasticity associated with inheritance of mitochondrial properties leads to higher transcription rate noise, and stochasticity in both mass and functional inheritance plays an important role in transcription rate noise. Contour lines on the bottom surface mark different values of $\eta_{\lambda}$. The `X' mark denotes the default parameterisation of our model. Other contour lines show that this relationship remains essentially identical when variability due to cell cycle stage and volume inheritance is removed, suggesting that $\sigma_n$ and $\sigma_f$ are the key sources of transcription rate noise.}
\label{transnoise}
\end{figure}

In Fig. \ref{transnoise}, the functional dependence of $\eta_{\lambda}$ on mitochondrial variability ($\sigma^2_n$ and $\sigma^2_f$) is shown from simulations. These results show that transcription rate noise is made up of significant contributions from both mitochondrial segregation and functionality. We also performed simulations where $\sigma_v$, the variability arising from uneven volume partitioning, was set to zero, and where cells were sampled at the same position in their cell cycle, removing different ages as a source of variability. As Fig. \ref{transnoise} shows, the removal of these sources of variability has little impact on the overall transcription rate noise level. These results lead us to conclude that mitochondrial sources of variability provide a strong contribution to cell-to-cell variability in transcription rate. This argument is supported by an approximate analytic treatment of the sources of error in transcription rate within our model (see `Estimating Noise Contributions' in Supplementary Information).

\subsection*{\label{gillmod}\iain{Mitochondrial variability can dominate noise in mRNA and protein expression}}
Having constructed and parameterised a model for mitochondrial variability and its effect on transcription in the cell, we now investigate the connection between these factors and downstream quantities: mRNA expression levels, and then (through further extension) protein expression levels. Noise in protein expression levels directly affects many cellular properties, as this noise causes cell-to-cell differences in the functional machinery available to perform cellular processes. Here we will investigate the influence of the mitochondrial variability suggested by the parameterisation of our model from experimental data on existing models for mRNA and protein expression. We connect our findings with the substantial existing body of literature on this topic in the Discussion section.

The production of mRNA and protein within a cell is often modelled using a master equation approach, addressing the probability of observing a given number of molecules at a given time. This analytical framework lends itself to the inclusion of our results for time-varying transcription rate (see Methods). Numerically, several studies have proposed techniques for incorporating time-varying rates in chemical kinetic systems \cite{jansen1995monte, haseltine2002approximate}: we use Shahrezaei \emph{et al.}'s modification \cite{shahrezaei2008colored} to the Gillespie simulation method \cite{gillespie1977exact} to simulate our model system. This protocol allows us to investigate the relative importance of intrinsic contributions (resulting in differences in expression levels between identical genes within a single cell) and extrinsic contributions (resulting in differences in expression between identical genes in different cells in a population).

\begin{figure}
\includegraphics[width=9cm]{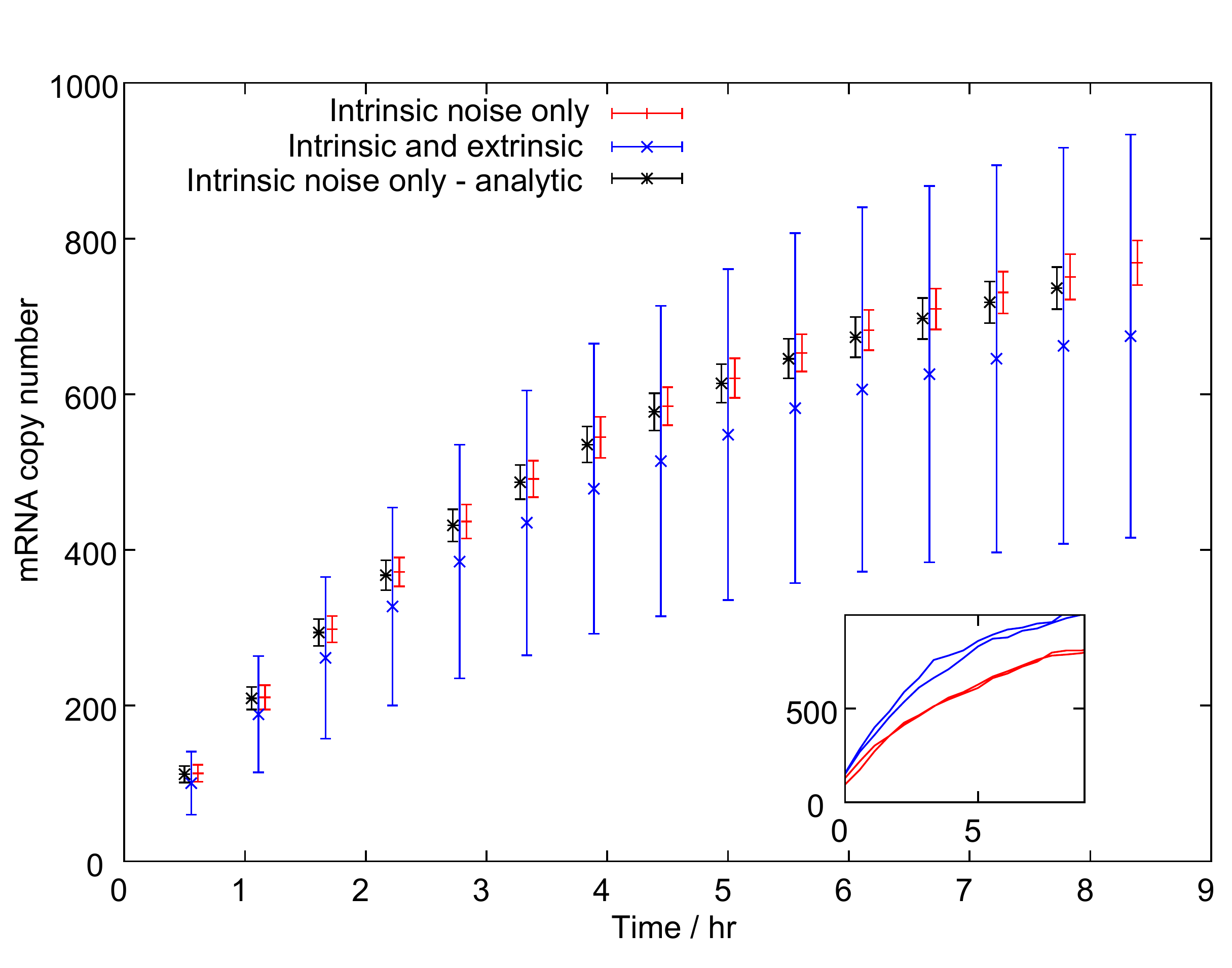}
\caption{\textbf{Mitochondrial variability contributes strongly to noise in mRNA levels.} Analytic and modified Gillespie simulation results for time evolution of mRNA levels with and without mitochondrial and volume variability. Bars show the mean and standard deviation of the corresponding distribution at a given time. Red ($+$) give simulated results without inherited variability. Black ($*$) give analytic results without inherited variability. Blue ($\times$) give simulated results with mitochondrial and volume variability, displaying much greater variance in mRNA expression. Bars are slightly offset in the x-direction for clarity. The inset shows two example time series for both simulated cases.}
\label{mrnalevels}
\end{figure}

Fig. \ref{mrnalevels} shows the increase in mRNA expression (from a level of zero at the start of the simulation) from our analytic approach incorporating changing transcription rate, and in simulations run using (see Methods) a parameter set from Raj \emph{et al.} \cite{raj2006stochastic}, in two scenarios: one involving only intrinsic noise effects (no noise due to mitochondrial variability) and one involving extrinsic noise in transcription rate due to mitochondrial mass, functionality, and cell volume variability, of the magnitudes found through parameterising our model with experimental data. It can be seen that mitochondrial variability leads to a large increase in the total noise in mRNA expression levels: without extrinsic factors, the noise in mRNA expression at a given time ($t = 8.3$ hours) was $\eta_m \simeq 0.04$, whereas $\eta_m \simeq 0.40$ with extrinsic factors.  We note that the means for the intrinsic and extrinsic noise cases differ: this result is due to the nonlinear dependence of transcription rate on ATP concentration, so that $[ATP]$ distributions with the same mean but different variances may yield transcription rate distributions with different means.



We can also perform simulations on the more complicated system involving protein production (see `mRNA \& Protein Levels' in Supplementary Information). With values from Raj \emph{et al.} \cite{raj2006stochastic} for protein degradation and translation rate (see Methods), this approach allows us to simulate dual reporter experiments, where the expression of two distinct but identically regulated protein-encoding genes is measured. Each protein was translated from a different mRNA strand, so these simulations tracked four quantities: the expression levels of the two mRNAs and the two proteins. Simulations were performed on synchronised and asynchronous cells, and with $\sigma_n, \sigma_f$ set to their model values and set to zero. In these simulations, mRNA molecules and proteins were also distributed binomially between daughter cells at mitosis (see Methods). 

\begin{figure}
 \includegraphics[width=9cm]{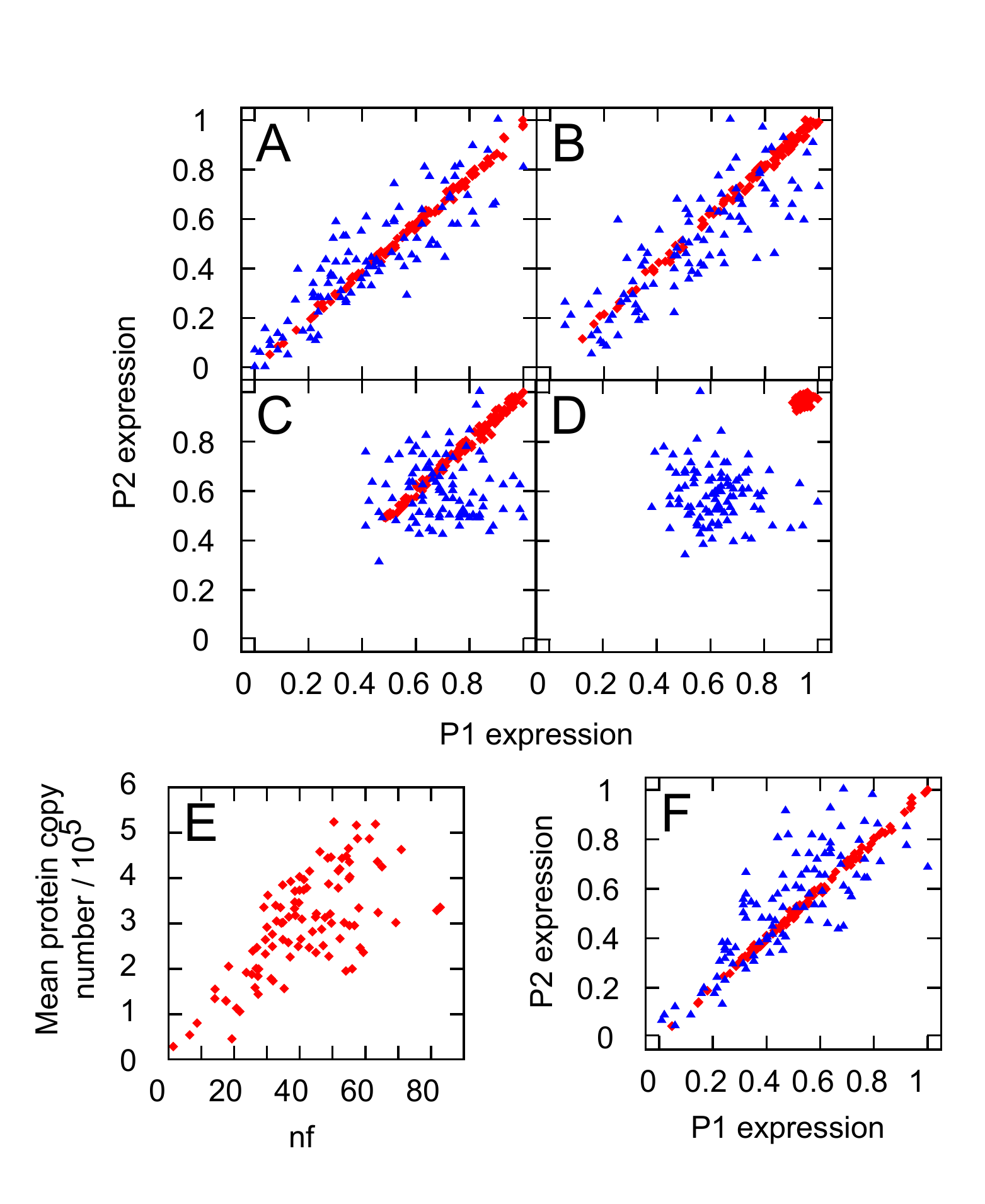}
\caption{\textbf{Effects of mitochondrial variability dominate protein expression variability in our model.} Dual reporter simulation with different sources of noise in our protein expression simulations. All plots except (E) are normalised so that the highest protein expression level in the cell population is 1. Red (diamonds) show results from Raj \emph{et al.}'s default parameterisation \cite{raj2006stochastic} used to model transcription, translation and degradation (see Methods). Blue (triangles) show results from this parameter set with degradation rates increased 100-fold. Protein levels are shown from population of (A) unsynchronised cells with mitochondrial and volume variability, (B) synchronised cells with mitochondrial and volume variability, (C) unsynchronised cells with no mitochondrial or volume variability, and (D) synchronised cells with no mitochondrial or volume variability. (E) Mean protein expression levels in the default parameterisation of Raj \emph{et al.} with the product of mitochondrial mass and function $nf$, in the system corresponding to (A). (F) The equivalent plot of (A) with translation rates independent of $[ATP]$.}
\label{dualrep}
\end{figure}

Dual reporter simulations performed with the parameterisation chosen from Raj \emph{et al.} \cite{raj2006stochastic} yield very low values for the magnitude of intrinsic noise. This low intrinsic noise was found to be due to the high copy number of proteins resulting from the parameterisation. To explore noise in systems with lower expression levels, we lowered the copy number of proteins by increasing the rates of mRNA and protein degradation (see Methods). Fig. \ref{dualrep} shows the resulting expression levels in two proteins with and without various sources of extrinsic noise, at the two different degradation rate protocols. These results show that, in our model, mitochondrial variability dominates the noise in protein expression levels. The spread of protein levels with mitochondrial and volume variability is much greater than the two-fold range achieved through cell cycle variability alone. Fig. \ref{dualrep} also illustrates that cells with higher mitochondrial mass and functionality generally have higher protein expression levels, though inheritance noise makes this correlation weaker. 

In our model, we find that energy variability arising through mitochondrial stochasticity is the dominant source of variability in transcription rate, mRNA and protein expression levels. However, we note that the causal factors of stochasticity in mRNA and protein levels within the cell are significantly more complicated than the simple transcriptional model presented above. The rates of many of the processes involved in more extended models are functions of many factors which our model does not include. The inclusion of these complicating terms rapidly makes an analytic description of the model impossible. However, we note that stochastic simulation techniques may be used to explore the behaviour of complex model given estimates for the functional dependence of process rates on extrinsic variables \cite{shahrezaei2008colored}.  

We also note that several studies have observed a decrease in intrinsic noise at higher levels of protein expression \cite{newman2006single, bareven2006noise}. We do observe such a decrease, though in the default parameterisation the magnitude of this effect is very small owing to the consistently low intrinsic noise levels.

\subsection*{\iain{Mitochondrial noise, by modulating transcription rate, can affect stem cell differentiation}}
\begin{figure}
\includegraphics[width=9cm]{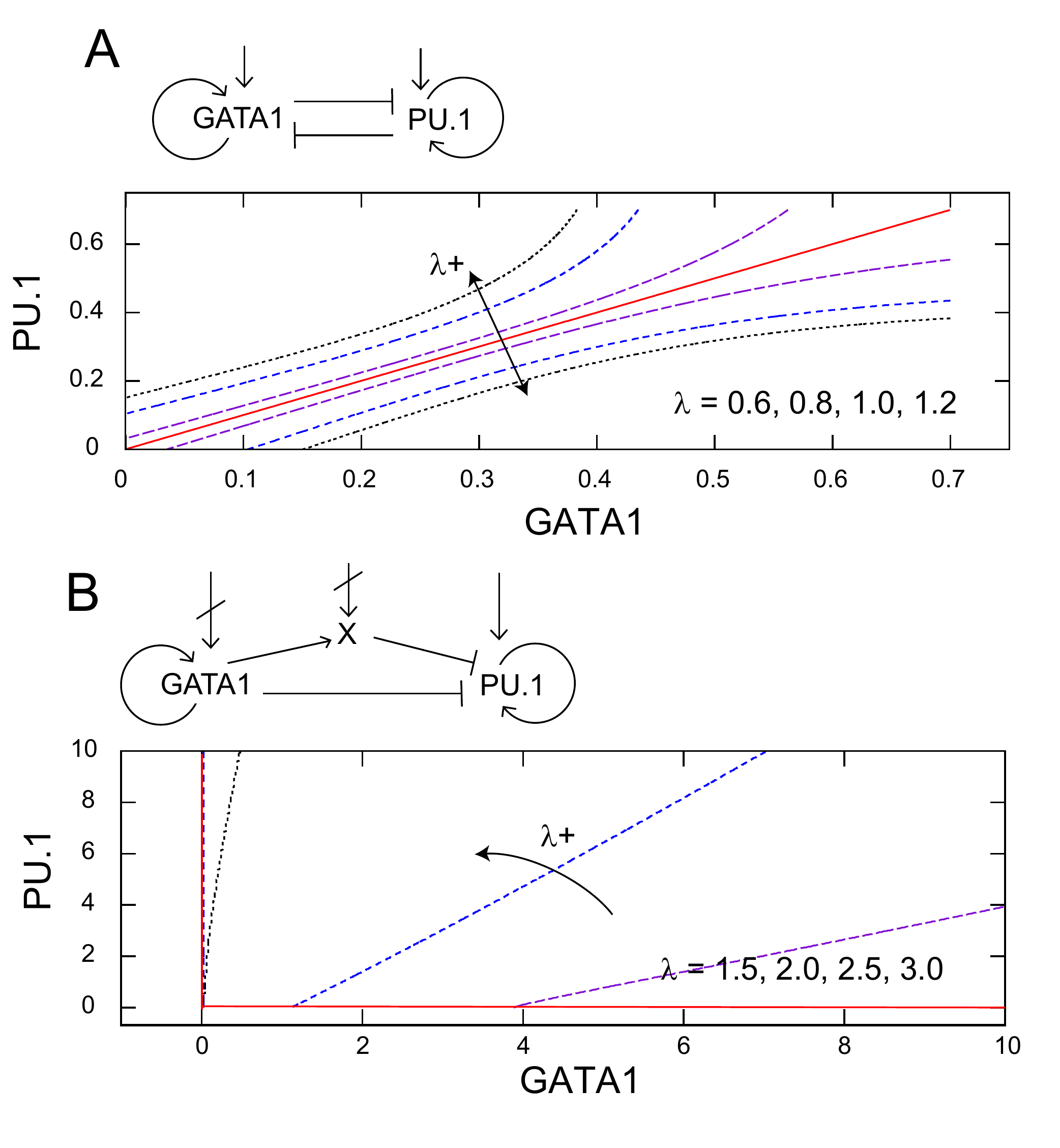}
\caption{\textbf{Transcription rate affects the stability of model stem cell systems.} In both diagrams, curves delineate the boundary of the attractor basin corresponding to the undifferentiated cell state. Red (solid) to black (dotted) lines show the basin structure as transcription rate $\lambda$ increases through the given values. (A) The structure of the undifferentiated attractor basin in the Huang model given different transcription parameters, showing the widening of the stable undifferentiated region at high transcription rate. (B) The structure of the undifferentiated attractor basin in the Chickarmane model, showing a decrease in undifferentiated basin size as transcription rate increases. \iain{The activation-repression structure of both models is illustrated -- in (B), external terms representing the activation of GATA1 and X exist but are set to zero in our analysis to allow PU.1 to be expressed under some conditions.} }
\label{n0props}
\end{figure}

As an illustrative application of our model, demonstrating its physiological relevance, we consider how,
through the extrinsic effects of [ATP] on protein levels, a link between mitochondrial content and stem
cell differentiation behaviour may arise. Differentiation dynamics in stem cells have often been modelled as the result of expression asymmetries in lineage regulation genes that interact in a regulatory network \cite{enver2009stem, macarthur2009systems, muller2008regulatory, graf2008heterogeneity}, but the initial sources of this expression variability have not been clearly elucidated and are a topic of active debate. Here we show that transcription rate variability resulting from mitochondrial variability can affect the dynamics of expression of such control genes. Experimentally, a link between stem cell differentiation and mitochondria was suggested by a recent study in mouse embryonic
stem cells \cite{schieke2008mitochondrial}, showing that pluripotent cells with low mitochondrial membrane potential had higher \emph{in
vitro} differentiation propensity, whereas those with higher membrane potential remained undifferentiated and formed large teratomas.

We explore two recent models for the cell fate decision between erythroid and
myeloid cell fates directed by the cross-antagonistic master lineage regulators GATA1 and PU.1. One model, by Huang \emph{et al.} \cite{huang2007bifurcation}, consists of a symmetric coupled ODE system for the expression levels of these two genes, including  cross-repression and self-activation term (see Methods). Another model, by Chickarmane \emph{et al.} \cite{chickarmane2009computational}, contains a similar but asymmetric ODE model, expanded to include interactions with a postulated third species which is promoted by GATA1 and represses PU.1. The Chickarmane \emph{et al.} model also includes external signalling terms which may act to promote GATA1 and PU.1, and repress the third species. In these models, cell states
are defined by the relative levels of expression of these genes, such that undifferentiated cells have comparable
levels of each transcription factor, while the two differentiated cell types correspond to a state with high levels
of one factor and low levels of the other. The interactions between the genes are parameterised by variables
such as self-activation and cross-repression rates (see Methods). The phase space for both these models comprises three
attractor basins, corresponding to the progenitor cell type and two differentiated cell types. 

Within the Huang model, at low protein expression levels, smaller perturbations are required to shift attractor basins than at high expression
levels -- a feature consistent across a large range of parameterisations. Varying the parameterisation of the model (modelling differentiation-inducing signalling) changes the
structure of these basins, so that the central undifferentiated basin becomes more or less stable to subsequent
perturbation. We vary the default parameterisation of the model in an attempt to assess the effect of changes
in transcription and translation rates (see Methods). We find that when the parameters related to the rate
of production of proteins are low, the central, undifferentiated state is less stable
than when they are high (see Fig. 7A), with a smaller volume of phase space leading to the undifferentiated basin.

Within the Chickarmane model, a different effect is observed. As before, we investigated the volume of phase space corresponding to the basin representing the undifferentiated state. We used a nonzero value for the external signalling term promoting PU.1 and explored the system at different transcription rates (see Methods).  We found that increasing the transcription rate led to a decrease in the range of values of the external interaction which supported a stable undifferentiated state (see Fig. 7B). This decrease in the stability of the undifferentiated state arose from a smaller volume of phase space leading to the undifferentiated basin as transcription rate increased, with more phase space occupied by the GATA1 basin. This result contrasts with the increased stability of the Huang model at high transcription rate, due to the importance of the third species (the expression of which is dependent on transcription rate): at high transcription rate, the increased strength of the combined effect of self-activating GATA1 and production of the third species shifts the basin structure strongly towards GATA1.

These results suggest that cell-to-cell variability in mitochondrial mass and function may, through induced variability in transcription rate, have a significant effect on the stability of bipotent cells. If differentiation dynamics are asymmetric and involve an intermediate species (as in the Chickarmane model), we find that high transcription rate destabilises the undifferentiated state. This destabilisation may be viewed as a result of the increased sensitivity of the system to perturbations: the asymmetric regulatory architecture means that a small increase in GATA1 will be quickly amplified at high transcription rate, as more GATA1 and X are quickly produced. If differentiation dynamics are symmetric and do not involve another species (as in the Huang model), high transcription rates increase the width of the basin corresponding to the undifferentiated state, acting to stabilise this state. This stabilisation is due to the increased robustness to perturbations afforded by the high production rate of both species at high transcription rate: without asymmetric interactions, the higher expression level of both genes makes the system less responsive to small perturbations. The results that emerge from this symmetric case gives results that are qualitatively comparable to an experimental study \cite{schieke2008mitochondrial} in which more cells with higher total mitochondrial membrane potential remained undifferentiated, suggesting that high mitochondrial performance stabilises the undifferentiated state.

Another, higher-order effect may conceivably play a role in both situations: several studies have found that, at high protein abundance levels (which may result from high transcription rates), intrinsic noise levels in protein expression decrease. While the
parameterisation of our dual reporter studies is such that these effects are small, the fact that less noise is
expected at higher protein expression levels suggests a third mechanism by which high mitochondrial content
may stabilise pluripotent cells. The contrasting results highlight the potential of experimental investigation of the effects of global transcription rate on the stability of multipotent states to inform of additional qualitative behaviors that models of lineage decision should be expected to exhibit.

\section*{Discussion}
We have introduced a crude mathematical model for the effects of stochasticity in mitochondrial segregation and functionality on transcription rate in cells. Our model, while simple enough to allow some analytic treatment, reproduces a good number of experimentally observed features concerning the interplay of mitochondrial properties and transcription rate. We analyse our model and find that mitochondria provide extrinsic noise contributions to transcription both through their uneven segregation at mitosis and through variability in their functionality.

We note that, in addition to requiring variability in the amount of mitochondrial mass, an adequate fit to our data required us to consider variability in the function of mitochondria. This connects with the wealth of recent experimental and theoretical interest regarding the causes and control of heterogeneity of mitochondrial function \cite{collins2002mitochondria, twig2008mitochondrial, kuznetsov2006mitochondrial, mouli2009frequency} and strengthens the case for the broad physiological relevance of functional variability.

We incorporate our results for mitochondrial-sourced extrinsic noise into existing models for mRNA and protein production, and show that mitochondrial noise can lead to significant variability between cells in a population. We also suggest that transcriptional variability resulting from mitochondrial noise may affect stem cell differentation, and illustrate this result with an analysis of two recent regulatory network-based models for stem cell differentiation. We find that the quantitative effect of transcription rate variability on stem cell differentiation depends on the architecture of the regulatory network under consideration.

Several recent studies have investigated the interplay between other possible sources of extrinsic noise in various organisms. Before concluding, we will discuss connections to this body of literature. The recent study by Huh and Paulsson \cite{huh2010non} found that variability in protein levels due to uneven inheritance at mitosis might explain a body of experimental data that was previously assumed to result from noise in the protein production process. A mathematical study by Rausenberger \emph{et al.} \cite{rausenberger2008quantifying} also investigated the effects of inheritance stochasticity on cellular noise. Our work bears significant parallels to these ideas, in that we postulate uneven inheritance of mitochondria to be a substantial contributing factor to noise in all cellular processes that require ATP, including the mechanisms of protein production. Our philosophy also mirrors part of the work of Huh and Paulsson in that our model considers a subset of cellular properties (in our case, mitochondrial partitioning and functionality, and cell volume) to provide all stochastic influences, with all other cellular properties evolving deterministically. 

The possible role of ATP as the proxy through which mitochondrial variability affects other cellular processes ties in with an early prediction of Raser and O'Shea \cite{raser2004control} who suggested that the dominance of extrinsic noise in expression variability across a wide range of proteins could result from fluctuations in a factor that affects expression for all genes. ATP, being required for the processes of transcription and translation, meets this criterion. Shahrezaei \emph{et al.} \cite{shahrezaei2008colored} illustrate the fact that extrinsic noise can influence intrinsic noise, through the former's effects on the rate constants involved in the latter. This influence plays an important role in our model, where extrinsic variability of mitochondrial properties influences the synthesis rates of mRNA and protein through their dependence on $[ATP]$. The ubiquity of ATP as an energy currency within the cell suggests that the rates of other intrinsic processes may be affected by the extrinsic variability we describe.

The link between the process of transcription and noise in protein expression levels that we explore in the last section of this paper is related to the findings of Blake \emph{et al.} \cite{blake2003noise} who found that protein expression noise depends on transcription efficiency. In our model, the modulation of transcription rate by noisy $[ATP]$ has downstream effects on protein noise levels.

Sigal \emph{et al.} \cite{sigal2006dynamic}, in a study of expression levels over a range of proteins, find cell cycle stage to be a significant contributor to extrinsic noise in protein abundance. Volfson \emph{et al.} \cite{volfson2005origins} used a mathematical framework to similarly identify population dynamics, and upstream transcription factors, as key extrinsic contributors to cellular noise. Our model is compatible with these results, as cells at different cell cycle stages will have had different protein expression histories over their lifetimes. However, we anticipate that mitochondrial variability will also provide a significant contribution to protein expression noise, through modulation of upstream processes.

An in-depth study by Newman \emph{et al.} in yeast cells \cite{newman2006single} found a variety of protein-specific differences in expression noise according to transcription mode and protein function. Our model does not capture protein expression noise in this level of detail. The study of Newman \emph{et al.} also characterised the contribution of intrinsic and extrinsic factors to total noise levels as a function of protein abundance. They found that while total expression noise did not scale with protein abundance, noise levels decreased with abundance when extrinsic factors were controlled for: suggesting that extrinsic factors were responsible for maintaining total noise levels as abundance increased. This suggestion that extrinsic noise increases in strength with protein abundance is captured in our protein level simulations. 

A study by Bar-Even \emph{et al.}, also in yeast cells \cite{bareven2006noise}, found intrinsic noise to be a substantial contributor to total noise, especially for proteins at intermediate abundance levels, with intrinsic contributions becoming less significant as expression levels increase (a similar result to Newman \emph{et al.}). In this and several of the other studies above \cite{newman2006single, blake2003noise}, fluctuations in mRNA number were postulated to be the most important source of noise in protein expression levels. One mRNA molecule may produce many copies of a protein, and regulatory and chromatin-remodelling influences result in stochastic production of mRNA molecules, so random `burstiness' in mRNA levels is a powerful source of noise in protein expression. 

Raj \emph{et al.} \cite{raj2006stochastic} studied noise in mRNA expression in detail, and identify intrinsic effects as the dominant factors. Their study found that genes located in close proximity to each other displayed synchronised expression, while the expression of genes that were physically separate was unsynchronised, suggesting that local rather than global effects determine the expression levels of genes. While this study demonstrated that intrinsic effects significantly contribute to total noise in some cases, it was not explicitly shown that the magnitude of these effects outweighed extrinsic effects. Our results are compatible with this view that intrinsic noise plays an important role in gene expression, but we suggest that extrinsic noise due to energy variability may also be an important contributor to overall noise levels.


We do not attempt to capture these mRNA processes explicitly: rather, we take transcription rate to be a function of $[ATP]$ as found in experiments \cite{neves2010connecting}. However, we note the result that the measured functional form of this relationship changes in experiments in which chromatin was decondensed. This result suggests that the functional form of transcription rate with $[ATP]$ allows us to capture some effects of the ATP-dependent chromatin remodelling process. 

\section*{Conclusions}
We find, through a phenomenological model constructed to reproduce recent data on mitochondrial and ATP variability, that stochastic inheritance of mitochondria at mitosis and variability in mitochondrial function may be important sources of noise in transcription. By extension, these factors may contribute significantly to noise in protein expression further downstream. We have proposed experimental tests to refine our model and demonstrate its application in existing models for mRNA and protein production and stem cell differentation, and discussed how these findings integrate into the current understanding of extrinsic noise in cellular biology. In particular, what our paper suggests is the need for multimodal single cell experiments through time (and through division) investigating coarse-grained measures of energy status, cellular volume, mitochondrial mass, and global rates of transcription and translation. Cellular variability is of central physiological importance but we suggest that to understand this we must elucidate the relationships between certain core variables, including the relationship between the machinery of expression and degradation and the energy status of the cell.

\section*{Methods}
\footnotesize
\textbf{Parameterisation of our model from experimental data.} We used a subset of available experimental data (shown in Fig. \ref{wufig1}) to choose numerical values for our key parameters. Some parameter values were fixed, in the sense that maximising the agreement between predictions from our model and a single experimental study allowed an optimal value to be chosen straightforwardly ($s_i, \sigma_v,$ and the ratio $\beta/\alpha$ were parameterised by data from Ref. \cite{neves2010connecting} (respectively the sigmoidal relationship between $[ATP]$ and transcription rate, the variance associated with volume partitioning, and the variance associated with mitochondrial partitioning), $v^*$ by the maximal cell volume observed in data from Ref. \cite{tzur2009cell}, and $\gamma$ by comparing the mean properties of simulated cells to mean ATP levels reported in a population of HeLa cells in Ref. \cite{wang1997turnover}). Other parameters influenced a range of predictions and values for these parameters were chosen by optimising the fit to experimental data across the set of results that they influenced (see `Fitting Other Parameters' in Supplementary Information).

Table \ref{paramvals} summarises the parameters and values employed in our model.


\begin{table*}

\begin{tabular}{|c|p{4cm}|p{4cm}|p{6cm}|}
\hline
Parameter & Description & Value & Motivation\\
\hline\hline
$f_a$ & $f$ memory term & 0.5 & Fit parameter -- chosen to give a mean functionality of 1 \\
$f_c^0$ & Sets mean functionality (control) & 0.5 & Fit parameter -- chosen to give a mean functionality of 1 \\
\hline
$v^*$ & Volume for mitosis (scale) & $2\,500\,\mu m^3$ & Fixed for consistency with maximum volume in Ref. \cite{tzur2009cell}  \\
$\gamma$ & Proportionality between $\frac{nf}{v}$ and $[ATP]$ & $39\,000\,\mu M \mu m^3$  & Fixed for consistency with mean ATP levels in Ref. \cite{wang1997turnover} \\
$\sigma_v$ & $v$ standard deviation at mitosis & $90\,\mu m^3$ & Fixed by volume segregation data in Ref. \cite{neves2010connecting} \\
$f^{(1,-1)}_c$ & Set mean functionality (with anti-oxidant and pro-oxidant respectively) & (0.69, 0.09) & Fixed by transcription rate noise levels in Ref. \cite{neves2010connecting}\\
$s_{1,2,3,4}$ & Fitting parameters for relationship between $[ATP]$ and $\lambda$ & 51.2, 44.7, $0.00288\,\mu$M$^{-1}$, $-1.9$ & Fixed by functional form of $\lambda$ in Ref. \cite{neves2010connecting} \\
\hline
$\alpha$ & $v$ growth rate & $0.92\,hr^{-1}$ & Chosen through optimisation -- constrained by mean cell cycle length in Ref. \cite{neves2010connecting}.\\
$\beta$ & $n$ growth rate & $0.022\,hr^{-1}$ & Fixed ratio with $\alpha$ through mitochondrial segregation data in Ref. \cite{neves2010connecting}\\
$\sigma_f$ & $f$ standard deviation at mitosis & 0.34  & Chosen through optimisation -- constrained through transcription rate noise and cell cycle length variability in Ref. \cite{neves2010connecting}\\
\hline
\end{tabular}
\caption{Parameters and values employed in our model. For further information see `Parameterisation of $\lambda(t)$' and `Fitting Other Parameters' in Supplementary Information.}
\label{paramvals}
\end{table*}

\textbf{Model Specifics.} In the following we provide details of our model. The key equations specifying the key dynamics of cells in our model are given by Eqns. \ref{firsteqn}-\ref{lasteqn}. The coupled ODEs in our model admit an analytic solution by simple integration:

\begin{eqnarray}
v(t) & = & v_0 + \frac{n_0 \alpha}{\beta}( \exp (\beta f t) - 1),\\
n(t) & = & n_0 \exp (\beta f t), 
\end{eqnarray}.

The reader will note that a variety of models for mitochondrial and volume growth will yield similar forms. In addition, other functional forms for the level of ATP may be suggested. A selection of these alternative forms are explored in `Other Models' in Supplementary Information. Within this model, at mitosis, 

\begin{eqnarray}
v_1 & = & \mathcal{N} \left( \frac{v^*}{2}, \sigma_v \right) \\
n_1 & = & \mathcal{N} \left( \frac{n}{2}, \sqrt{\frac{n}{4}} \right), 
\end{eqnarray}

where $\mathcal{N} (\mu, \sigma)$ is a normal distribution with mean $\mu$ and variance $\sigma^2$, and $v_2 = v-v_1, n_2 = n-n_1$ determine the daughter volumes $v_1, v_2$ and mitochondrial masses $n_1, n_2$. The variances of the mitochondrial distribution is chosen to represent a binomial distribution ($\mathcal{B}(n, p) \simeq \mathcal{N}(np, \sqrt{np(1-p)})$, with $p = \frac{1}{2}$, for high $n$). The variance of the volume distribution is chosen to match experimental data on volume partitioning.

Functionality evolves through changes at mitosis events, which follow an AR(1) process. A daughter cell's functionality $f^D$ is determined from its parent's functionality $f^P$:

\begin{equation}
f^D = f_a f^P + f_c + \mathcal{N}(0, \sigma_f)
\label{fdist}
\end{equation}

Upon mitosis, both daughter cells inherit the same $f^D$ value, drawn from Eqn. \ref{fdist}. Once chosen, a cell's functionality remains constant throughout one cell cycle. We can model treatment with anti- and pro-oxidants by respectively increasing and decreasing $f_c$, raising or lowering the mean functionality of mitochondria. With $\sigma_f = 0.3$, there is a very small but finite chance that cells will inherit zero (or lower) functionality. To avoid this unphysical case, we impose a cutoff of $0.01$ on mitochondrial functionality. Values under this cutoff are resampled from Eqn. \ref{fdist}. Another model, where $f$ varies continuously within a cell cycle, is considered in `Other Models' in Supplementary Information.

\textbf{Virtual Mitochondria.} In our model, the standard deviation in mitochondrial mass at mitosis is a function of the number of virtual mitochondria in the cell, $\sigma_n = \sqrt{\frac{n}{4}}$, from binomial partitioning. We can vary the number of virtual mitochondria in the cell while keeping the total mitochondrial volume constant, by changing the volume assigned to an individual virtual mitochondrion. Let an individual virtual mitochondria have volume $\frac{v_n}{N}$, and let there be $N$ virtual mitochondria in a cell. The total mitochondrial volume is $v_n$, and the mean inherited mitochondrial volume is half of this. The standard deviation in virtual mitochondrial number from a binomial distribution of the virtual mitochondria is $\sqrt{\frac{N}{4}}$, so the standard deviation of mitochondrial volume is $\frac{2 v_n}{\sqrt{N}}$. A standard deviation of $\sigma_n$ then corresponds to $N = \frac{4v_n^2}{\sigma_n^2}$ virtual mitochondria per cell.

\textbf{Cell Dynamics Simulations.} To simulate a population of cells, we used a simple Euler method to solve the dynamic equations. A population of $N = 10\,000$ cells was simulated. When mitosis occurred, a random cell from the population was chosen for replacement by the new cell. To measure distributions, this procedure was continued until the distributions stabilised. To measure sister-sister and between-generation correlations, a list of relationships between cells was maintained, with sister pairs only sampled when both sisters underwent an entire cell cycle without replacement. We note that this removal of randomly-chosen cells may potentially introduce artefacts into the results, as the constant probability of cell removal means that the probability of a cell surviving to a certain age is decreasing. We checked our results with a different simulation protocol: running the system and allowing exponential growth up to $10^6$ cells, with no removal. The resulting distributions were indistinguishable from the first protocol, showing that the removal rate is low enough so that the statistics are not affected.

\textbf{mRNA \& Protein Expression Simulations.} For the simple system illustrating mRNA expression levels alone, cells were simulated using the modification of Shahrezaei \emph{et al.}'s \cite{shahrezaei2008colored} to the Gillespie simulation method \cite{gillespie1977exact} in two scenarios: one involving only intrinsic noise ($\sigma_n = \sigma_v = \sigma_f = 0$) and one involving extrinsic noise due to mitochondrial mass, functionality, and cell volume variability. In both cases, the initial copy number of mRNA molecules was set to zero, to illustrate differences in the dynamics of the system: transcription started at the birth of the cell, and there were no effects from mRNA inheritance. In the intrinsic noise experiments, a population of cells were simulated from identical initial conditions, with their $n, v, f$ values set to the means of those variables obtained from simulations. The variability in mRNA expression levels in these cells was therefore solely due to intrinsic noise. For the extrinsic noise simulations, mRNA expression was simulated in the heterogeneous population of cells that resulted from dynamic simulation. An ensemble of $2\,500$ cells was analysed for both cases and the mRNA content at each timestep recorded.

To simulate the more complicated systems and investigate inheritance effects, we coupled the Shahrezaei \emph{et al.} simulation protocol with the simple ODE solver so that simulation of a population of cells growing and producing mRNA and protein involved the following algorithm. 1) Use the ODE solver to calculate a cell's time of mitosis and the time series of volume and transcription rate throughout the cell lifetime. 2) Use Shahrezaei \emph{et al.}'s method to compute the time behaviour of mRNA and protein levels given these time series for production rates. 3) Create daughter cells with noisy partitioning of volume, mitochondrial mass (binomial) and functionality (AR(1)), and mRNA and protein copy numbers (binomial).

After Raj \emph{et al.} \cite{raj2006stochastic}, we employ the following model values for birth ($\lambda$) and death ($\xi$) rates of mRNA ($m$) and protein ($n$): $\langle \lambda_m \rangle = 0.06\,s^{-1}, \langle \lambda_n \rangle = 0.007\,s^{-1}, \xi_m = 7 \times 10^{-5}\,s^{-1}, \xi_n = 1.1 \times 10^{-5}\,s^{-1}$. In the case of birth rates, we used these mean values to scale the $\lambda([ATP])$ curves from das Neves \emph{et al.} \cite{neves2010connecting} so that the mean $[ATP]$ level observed in a population gave the mean $\langle \lambda \rangle$ values from Raj \emph{et al.} (see `mRNA \& Protein Levels' in Supplementary Information). We also ran experiments with the degradation rates increased 100-fold: $\xi_m = 7 \times 10^{-3}\,s^{-1}, \xi_n = 1.1 \times 10^{-3}\,s^{-1}$, to explore the behaviour of the system at lower expression levels.

\textbf{Experiments.} CMXRos labelling was done according to manufacturer (invitrogen) instructions, cells 
were incubated with the probe (75 nM) for 15 min. at 37ºC and washed with warm PBS.

For the dual dye experiments, cells were loaded simultaneously with CMXRos and 
MitoTracker Green FM dye (Molecular Probes) for 20 min and after a brief PBS rinse 
fixed in PBS with 4\% paraformaldehyde.

Cell cycle length experiments were done by growing cells in the presence of media 
alone or containing either the anti-oxidant Dithiothreitol (250 microM DTT) or the 
pro-oxidant Diamide (50 microM). Cell cycle length was measured as the time interval 
between two mitotic events in a single cell, analysed by live cell imaging using the 
Cell IQ platform and image analysis software (Chipman Tech.).

In the flow cytometry experiments, tripsinised Hela cells were washed and incubated with 20 nM MitoTracker Green FM dye (Molecular Probes) in medium at 37 C for 15 min, and then washed. Cells were analyzed using flow cytometry (Dako CyAn ADP).

\textbf{Master equations.} Let us consider the master equation for the transcription process, describing the probability of observing the system with $m$ mRNAs at time $t$:

\begin{equation}
\frac{\partial P_m}{\partial t} = \lambda(t) P_{m-1} + \xi (m+1) P_{m+1} - (\lambda(t) + \xi m) P_m,
\end{equation}

where $P_m$ is the probability of observing $m$ mRNAs at a given time $t$, $\lambda(t)$ is transcription rate, and $\xi$ is an mRNA degradation rate. 

Using a linear approximation for $\lambda(t)$ (see `Parameterisation of $\lambda(t)$' in Supplementary Information), we can solve this by using a generating function approach (see `mRNA \& Protein Levels' in Supplementary Information). The mean, variance, and probability distribution of mRNA copy number at arbitrary time are given by:

\begin{eqnarray}
\mu_m & = & (a_3 + 1)^{m_0 - 1} e^{a_1 + a_2} (a_1 + a_1 a_3 + m_0) \\
\sigma^2_m & = & (a_3 + 1)^{m_0} e^{a_1 + a_2} \left( a_1 + a_1^2 + \frac{m_0}{a_3 + 1} \right. \nonumber  \\
&& \left. \times \left( 1 + 2 a_1 + \frac{m_0 - 1}{a_3 + 1} \right) - (a_3 + 1)^{m_0 + 2} \right. \nonumber \\ 
&& \bigg. \times e^{a_1 + a_2} (a_1 + a_1 a_3 + m_0)^2 \bigg) \\
P_m & = & a_3^{m_0-m} e^{a_2} \frac{m_0!}{m! (m_0 - m)!} {}_1F_1(-m; m_0 - m + 1; -a_1 a_3)
\end{eqnarray}

where ${}_1F_1$ is the Kummer confluent hypergeometric function, $a_1 =  \frac{1}{\xi} \left(c + bt - c e^{-\xi t} + \frac{b}{\xi} (e^{-\xi t} - 1) \right) 
,\,a_2 = -m_0 t \xi - a_1$ and $a_3 = e^{\xi t} - 1$.

\textbf{Stem Cell Model.} The progenitor cell differentation model of Huang \emph{et al.} \cite{huang2007bifurcation} consists of the following equations for the evolution of protein expression levels $x_1$ (GATA1) and $x_2$ (PU.1):

\begin{eqnarray}
\frac{d x_1}{dt} & = & a_1 \frac{x_1^n}{\theta_{a1}^n + x_1^n} + b_1 \frac{\theta_{b1}^n}{\theta_{b1}^n + x_2^n} - k_1 x_1 \\
\frac{d x_2}{dt} & = & a_2 \frac{x_2^n}{\theta_{a2}^n + x_2^n} + b_2 \frac{\theta_{b2}^n}{\theta_{b2}^n + x_1^n} - k_2 x_2 
\end{eqnarray}

In this parameter set, $a$ variables are self-activation rates, $b$ variables are cross-repression rates, $k$ are decay rates, and $\theta$ and $n$ control the functional form of these processes. Huang \emph{et al.} show that altering these parameters changes the structure of the corresponding attractor landscape, so that the central undifferentiated attractor basin changes in size, affecting the predisposition of the system to differentiate. This landscape change gives a `priming' of the system such that the effect of subsequent asymmetries may vary. In this study, we vary the landscape by symmetrically varying $a$ and $b$, the parameters associated with activation and repression, from the default parameterisation $a_i = b_i = k_i = 1, n = 4, \theta_{ai} = \theta_{bi} = 0.5$ for $i = 1,2$. Specifically, we modulated $a$ and $b$ with a multiplicative factor $\lambda$: $a \rightarrow \lambda a, b \rightarrow \lambda b$. We draw the connection between higher values of $a$ and $b$ and higher transcription and translation rates, as the rate of production of chemical species is increased by an increase in these parameters.

The model of Chickarmane \emph{et al.} \cite{chickarmane2009computational} involves a relationshup between three dynamic variables:

\begin{eqnarray}
\frac{d [G]}{dt} &=& \frac{\alpha_1 A + \alpha_2 [G]}{1 + \beta_1 A + \beta_2 [G] + \beta_3 [G][P]} - \gamma_1 [G] \\
\frac{d [P]}{dt} &=& \frac{\delta_1 B + \delta_2 [P]}{1 + \epsilon_1 B + \epsilon_2 [P] + \epsilon_3 [G][P] + \epsilon_4 [G][X]} - \gamma_2 [P] \\
\frac{d [X]}{dt} &=& \frac{\zeta_1 [G]}{1+ \eta_1 [G] + \eta_2 C} - \gamma_3 [X]
\end{eqnarray}

where $[G], [P], [X]$ are the concentrations of GATA1, PU.1 and a postulated chemical species X respectively. $A, B, C$ are external signalling factors: $A$ and $B$ promote GATA1 and PU.1 respectively, and $C$ represses X. Other variables take default values $\alpha_1 = \beta_1 = \delta_1 = \epsilon_1 = \beta_3 = \epsilon_3 = 1$, $\alpha_2 = \beta_2 = \delta_2 = \epsilon_2 = 0.25$, $\epsilon_4 = 0.13$, $\gamma_1 = \gamma_2 = \gamma_3 = 0.01$, $\zeta_1 = \eta_1 = 0.01$, $\eta_2 = 10$. To vary transcription rates, we modulate $\alpha, \delta$ and $\zeta$ terms with a multiplicative factor which we take to be proportional to transcription rate $\lambda$. We used $B = 0.5$ with $A = C = 0$, allowing an external signal that promotes PU.1, to promote stability of the undifferentiated state. 


\normalsize

\section*{Acknowledgements}
The authors wish to thank Sumeet Agarwal, Maria Domingo-Sananes, Bela Novak and Vahid Shahrezaei for their valuable comments. IGJ and NSJ acknowledge funding from the BBSRC (Grant number: BB/D020190/1), and BG and NSJ acknowledge funding from the EPSRC. FJI has been funded by the Ministerio de Ciencia e Innovacion (Grant number: BFU2009-10792). RPdN is thankful to CNC and MIT-Portugal Program for support.

\onecolumngrid 

\section*{Supplementary Information}

\subsection{\label{nuparam} Parameterisation of $\lambda(t)$}
\begin{figure}
\includegraphics[width=8cm]{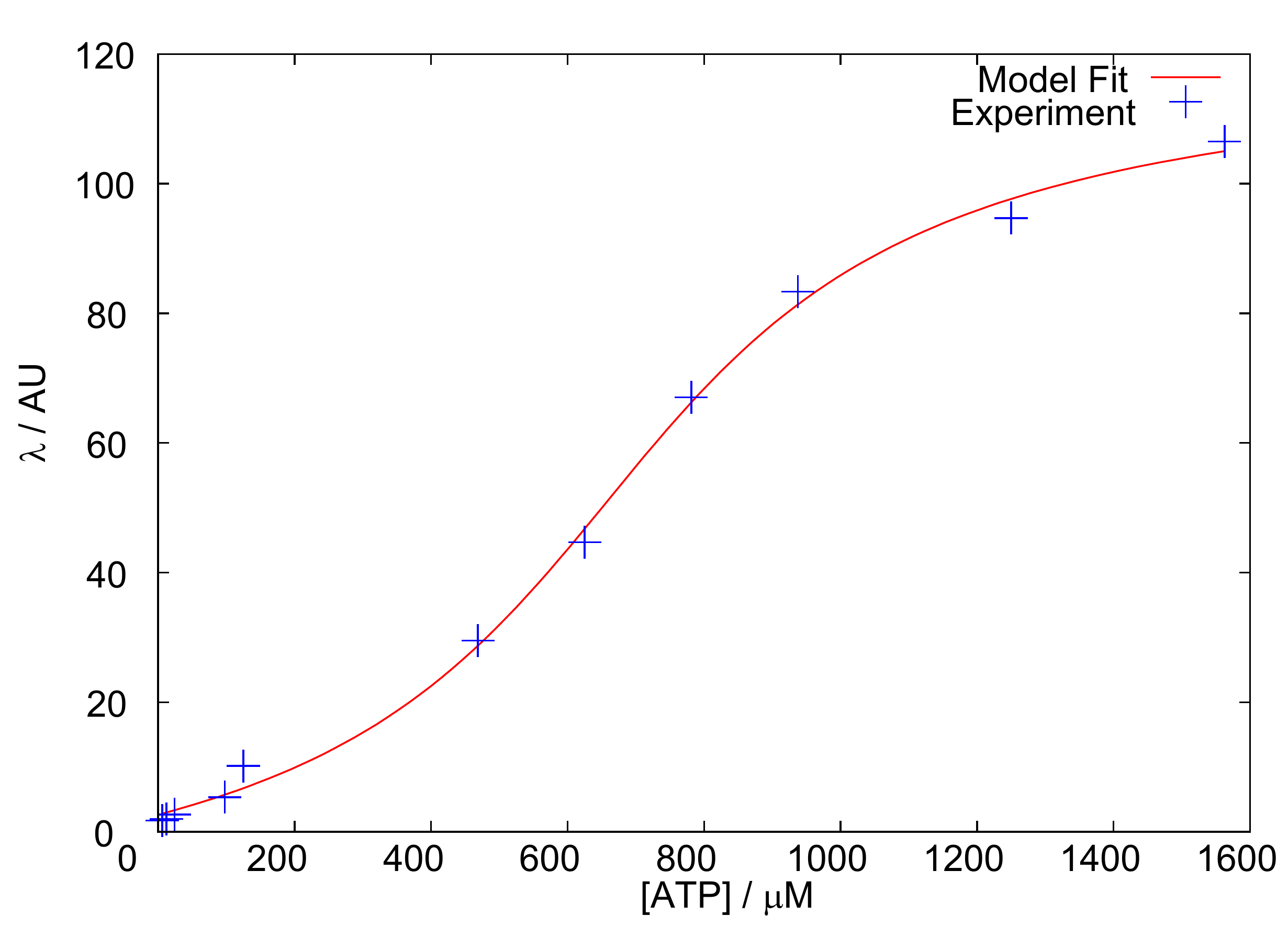}
\caption{$\lambda$ with $[ATP]$ (experiment and fit).}
\label{sigmoid}
\end{figure}

Figs. 2g and 2i in das Neves \emph{et al.} \cite{neves2010connecting} give the response of transcription rate ($\lambda$) in arbitrary units to varying concentrations of ATP, without and with artificial decondensation of chromatin respectively. Measuring $[ATP]$ in $\mu M$ and working with the same arbitrary units employed in that study, we model this response with the expression:

\begin{equation}
\lambda = s_1 + s_2 \tan^{-1} (s_3 [ATP] + s_4 ),
\label{nuatp}
\end{equation}

with $s_1 \simeq 51.2, s_2 \simeq 44.7, s_3 \simeq 2.88 \times 10^{-3} \mu M^{-1}, s_4 \simeq -1.9$ (see Fig. \ref{sigmoid}). 

das Neves \emph{et al.} also produced a curve showing $\lambda$ with $[ATP]$ in an experimental situation involving the decondensation of chromatin in the cell. This curve can be parameterised by the above equation with $s^d_1 \simeq 40.0, s^d_2 \simeq 54.7, s^d_3 \simeq 1.6 \times 10^{-3} \mu M^{-1}, s^d_4 \simeq -0.27$.

We choose this inverse tangent functional form to model the response of transcription rate to $[ATP]$ for mathematical simplicity in subsequent sections, and note that modelling with other functional forms (for example, Hill functions) is also possible.

In the parameterisation of our model, the time series of $[ATP]$ in cells is rather linear and slowly varying. This behaviour emerges both due to the dynamics of mitochondrial density (which tends towards $\frac{\beta}{\alpha}$ with time) and the fact that the exponential growths involved are cell-cycle limited to only span a factor of around two before mitosis. It will be useful to employ a linear approximation to $\lambda(t)$, gained from an expansion in $t$ about $t'$:

\begin{eqnarray}
\lambda(t) & = & s_1 + s_2 \tan^{-1} \left( s_4 + s_3 \frac{\gamma f n_0 e^{\beta f t}}{v_0 + \frac{n_0 \alpha}{\beta} ( e^{\beta f t} - 1)} \right) \\
& \simeq & c + bt,
\end{eqnarray}

where the values of $c$ and $b$ are found after some algebra to be:

\begin{eqnarray}
b & = & \frac{s_2 f \beta t' d}{(1 + (s_4 + d)^2)} \\
c & = & s_1 + s_2 \tan^{-1} (s_4 + d) - b,
\end{eqnarray}

and

\begin{equation}
d = \frac{s_3 f n_0 \beta \gamma e^{\beta f t'}}{\alpha n_0 \left( e^{\beta f t'} - 1 \right)  + \beta v_0 }.
\end{equation}




Constants $c$ and $b$ then depend on cellular initial conditions and $t'$ (which we take to be half the mean cell cycle length) but not on $t$. The sign of $b$ is determined by the over- or under-population of mitochondria in the cell at mitosis: over-population will lead to high mitochondrial density and mean-reversion will act to decrease transcription rate with time (and \emph{vice versa}).

\subsection{Fitting Other Parameters}

The parameters in our model concerning volume cutoff and partitioning were chosen straightforwardly: $v^*$ for consistency with Ref. \cite{tzur2009cell} and $\sigma_v$ to best fit experimental data from Ref. \cite{neves2010connecting}. We then needed to find values for the remaining parameters influencing properties of our model cells, namely $\alpha$ and $\beta$, the growth rates of cell volume and mitochondrial mass, $\gamma$, the constant of proportionality between $\frac{nf}{v}$ and $[ATP]$, and parameters determining the statistics of mitochondrial functionality $f_a, f_c^0, \sigma_f$. To fix values for these parameters, we first chose values for two parameters ($f_a$ and $f_c^0$) to give a mean functionality of 1 in the control population, for simplicity. An optimisation procedure was then used to select values for $\sigma_f$, $\alpha$, $\beta$ and $\gamma$. The procedure we used involved choosing arbitrary initial conditions for each parameter (though the order of magnitude of these initial values was determined by a crude preliminary investigation) and iterating, choosing a new value for one of the free parameters at each step, and retaining this value if the overall performance of the parameter set improved. The new values were chosen either (with probability 0.1) uniformly over the order of magnitude associated with that parameter or (with probability 0.9) uniformly from the interval of $10\%$ deviations from the old value. To judge performance, a score for each parameter set was calculated based on the absolute deviation between experimental and simulated observables, averaged over the seven quantities shown in Table \ref{paramfit}. Once these values were chosen for cells under `normal' oxidative conditions (the default), $f_c^{-1, 1}$ were set to match the experimentally observed transcription noise levels under the corresponding oxidative conditions.

\begin{table}
\begin{tabular}{l|l|l|l}
Observable & Experimental Value & Simulated Value & Error \\
\hline
$\eta_{\lambda}$ Control & 0.4 & 0.389 & 0.028 \\
$\eta_{\lambda}$ Anti-oxidant & 0.2 & 0.204 & 0.019 \\
$\eta_{\lambda}$ Pro-oxidant & 1 & 0.943 & 0.057 \\
$\eta_{\lambda}$ Sister Cells & 0.08 & 0.093 & 0.166 \\
SD $n_+/n_-$ & 0.23 & 0.217 & 0.055 \\
Cell cycle length & 30.5 & 30.8 & 0.009 \\
Mean $[ATP]$ & 900 & 925 & 0.028 \\
\end{tabular}
\caption{Experimental observables and simulated fits obtained through optimisation of fitting parameters. Errors are the absolute difference between simulated and experimental values divided by experimental value. The score for a parameter set is the sum of errors for each observable.}
\label{paramfit}
\end{table}

\subsection{Mitochondrial Membrane Potential}
\begin{figure*}
\includegraphics[width=13cm]{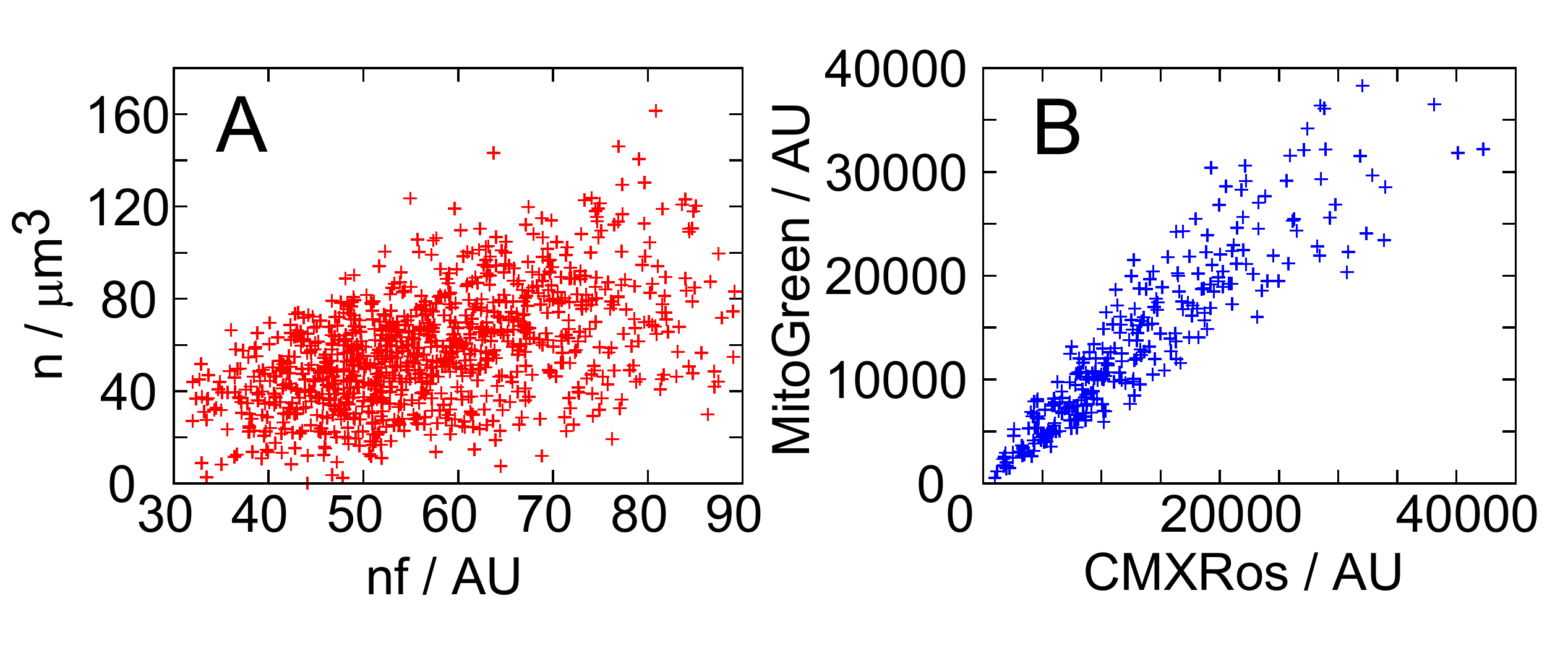} 
\caption{(A) Simulated relationship between a measure of total membrane potential in a cell (the product of mitochondrial mass $n$ and functionality $f$) and the mitochondrial mass in the cell. (B) Membrane potential (measured with CMXRos) against mitochondrial content (measured with MitoGreen).}
\label{dualdye}
\end{figure*}

Fig. \ref{dualdye} shows a comparison of the relationship between total membrane potential in a cell and mitochondrial mass in the cell, from new experiments (see Methods in Main Text) and simulation of our model. Both our model and experimental data shows a linear correlation between total membrane potential and mitochondrial mass. This result emerges straightforwardly from our model due to the representation of total membrane potential (the product of $n$ and $f$) and the qualitative agreement with experiment suggests that this modelling approach is suitable.

\subsection{\label{pdist} Estimating Noise Contributions}
Here we find an expression for transcription rate in terms of the initial conditions and age of a cell, then estimate the variances associated with each of these contributory factors in order to assess the strength of each contribution to overall transcription rate noise.

As cells evolve deterministically within a cell cycle, the concentration of ATP in a model cell is a function of the cell's initial conditions and its age:
\begin{eqnarray}
[ATP] & = & \frac{\gamma f n}{v} \\
& = & \frac{\gamma f n_0 e^{\beta f t}}{v_0 + \frac{n_0 \alpha}{\beta} (e^{\beta f t} - 1)},
\end{eqnarray}

where the second line follows from simple integration of our dynamic equations. If we assume that the initial conditions of an individual cell are uncorrelated -- which is the case for single divisions within our model, but population and memory effects may possibly lead to correlations -- we can use standard rules for combining means and variances to calculate the population mean and variance of the ATP distribution from the total differential:

\begin{eqnarray}
\mu_{ATP} & = & \frac{\gamma \mu_f \mu_{n_0} e^{\beta \mu_f \mu_t}}{\mu_{v_0} + \frac{\mu_{n_0} \alpha}{\beta} (e^{\beta \mu_f \mu_t} - 1)}.\\
\sigma^2_{ATP} & = & \left| \frac{\partial ATP}{\partial n_0} \right|^2 \sigma^2_{n_0} + \left| \frac{\partial ATP}{\partial f} \right|^2 \sigma^2_f + \left| \frac{\partial ATP}{\partial v_0} \right| \sigma_{v_0}^2 + \left| \frac{\partial ATP}{\partial t} \right|^2 \sigma^2_t \label{etanugen}\\
& = & \frac{e^{2 f t \beta } \beta ^2 \gamma ^2}{\left(\left(e^{f t \beta } - 1\right) n_0 \alpha +v_0 \beta \right)^4} \bigg(  n_0^2 \left(n_0 \alpha  \left(e^{f t \beta }-1-f t \beta \right)+v_0 \beta  (1+f t \beta )\right)^2 \sigma_f^2  \nonumber \\
&& + f^2 v_0^2 \beta ^2 \sigma_{n_0}^2+f^4 n_0^2 \beta ^2 (n_0 \alpha -v_0 \beta )^2 \sigma_t^2+f^2 n_0^2 \beta ^2 \sigma_{v_0}^2 \bigg) \\
& = & W_{n_0}^2 \sigma_{n_0}^2 + W_{v_0}^2 \sigma_{v_0}^2 + W_{t}^2 \sigma_t^2 + W_{f}^2 \sigma_{f}^2.
\end{eqnarray}


As transcription rate $\lambda$ depends solely on $[ATP]$ in our model, we can then calculate the mean and variance of the distribution of $\lambda$:

\begin{eqnarray}
\mu_{\lambda} & = & s_1 + s_2 \tan^{-1} (s_3 \mu_{ATP} + s_4) \\
\sigma^2_{\lambda} & = & \left| \frac{\partial \lambda}{\partial ATP} \right|^2 \sigma^2_{ATP} \\
& = & \left( \frac{s_2 s_3 \sigma_{ATP}}{1 + (\mu_{ATP} s_3 + s_4)^2} \right)^2.
\end{eqnarray}

From this, we can explore the contributions of mitochondrial segregation ($\sigma^2_{n_0}$) and functional diversity ($\sigma^2_f$) on transcription rate noise $\eta_{\lambda} = \frac{\sigma_{\lambda}}{\mu_{\lambda}}$. The overall expression can be written:

\begin{eqnarray}
\eta_{\lambda} & = & \frac{\sigma_{\lambda}}{\mu_{\lambda}} \\
& = & \frac{s_2 s_3 \sigma_{ATP}}{(s_1 + s_2 \tan^{-1} (s_3 \mu_{ATP} + s_4)) (1 + (\mu_{ATP} s_3 + s_4)^2)} \\
& = & \frac{s_2 s_3 \sqrt{W_{n_0}^2 \sigma_{n_0}^2 + W_{v_0}^2 \sigma_{v_0}^2 + W_{t}^2 \sigma_t^2 + W_{f}^2 \sigma_{f}^2 }}{(s_1 + s_2 \tan^{-1} (s_3 \mu_{ATP} + s_4)) (1 + (\mu_{ATP} s_3 + s_4))^2}  \\
& = & w \sqrt{w_f \eta^2_f + w_{n_0} \eta^2_{n_0} + w_t \eta^2_t + w^2_{v_0} \eta_{v_0}}, \label{sqrtcontrib},
\end{eqnarray}

where the last line condenses the expression into the quadrature sum of noise levels in cellular parameters with weighting factors $w_i = W^2_i \mu^2_i$.  

\textbf{Initial volume distribution.} The mean and variance of initial volumes is simple to obtain, as division always occurs at a fixed volume and proceeds according to a fixed distribution. We straightforwardly have $\mu_{v_0} = \frac{v^*}{2}, \sigma_{v_0} = \sigma_v$.

\textbf{Initial mitochondrial mass distribution.} We estimate the population variance of initial mitochondrial mass as the variance associated with the daughter of a cell with population average properties just before mitosis. Such a cell has $v = v^*, n = \frac{\beta}{\alpha} v^*$. The mean and variance associated with initial mitochondrial mass in the daughter of such a cell is $\mu_{n_0} = \frac{\beta}{\alpha}\frac{v^*}{2}, \sigma_{n_0} = \sqrt{\frac{\beta v*}{4 \alpha}}$. These values give a simple estimate comparing to numerical results on the population statistics (see Fig. \ref{fdist}A).

\textbf{Mitochondrial functionality distribution.} Similarly, we estimate the population distribution of mitochondrial functionalities by considering the daughter of a population-average cell. The steady-state distribution of the AR(1) process determining $f$ is known to be normal with mean $\frac{f_c}{1-f_a}$ and variance $\frac{\sigma_f^2}{1 - f_a^2}$. It is not straightforwardly obvious that the distribution of $f$ in an unsynchronised population of cells, where the lifespan of a cell is a function of $f$, should also follow this form, but numerical results confirm that the $f$ distribution does closely match it (see Fig. \ref{fdist}B).

\textbf{Distribution of cellular ages.} This distribution is hard to estimate or approach analytically. We obtained the required values numerically from simulations, as shown in Fig. \ref{timedist}. The mean and variance of this distribution was found to vary with $\sigma_{n_0}$ and $\sigma_f$, the variance in mitochondrial mass and mitochondrial function. Over the ranges $0 < \sigma_{n_0} < 20$ and $0 < \sigma_f < 0.4$, $\mu_t$ varied between $10$ and $25$ hours and $\sigma_t$ varied between $9$ and $50$ hours. However, these terms provide a negligible contribution to the transcription rate variability (see later), so we approximate the mean and variance of $P(t)$ to be constant, with $\mu_t \simeq 16$ and $\sigma_t \simeq 14$, the values corresponding to the default model parameterisation.

We find, with our default parameter set, $w_f = 8.5 \times 10^5, w_{n_0} = w_{v_0} = 4.2 \times 10^5$, and that $w_t$ is zero within working precision. Typical noise levels in these quantities were measured from simulations as $\eta_f = 0.34, \eta_{n_0} = 0.13, \eta_{v_0} = 0.07$. This approximate analytic approach thus supports the hypothesis that the dominant contributions to transcription rate variability are from mitochondrial variability terms.

\begin{figure*}

\includegraphics[width=15cm]{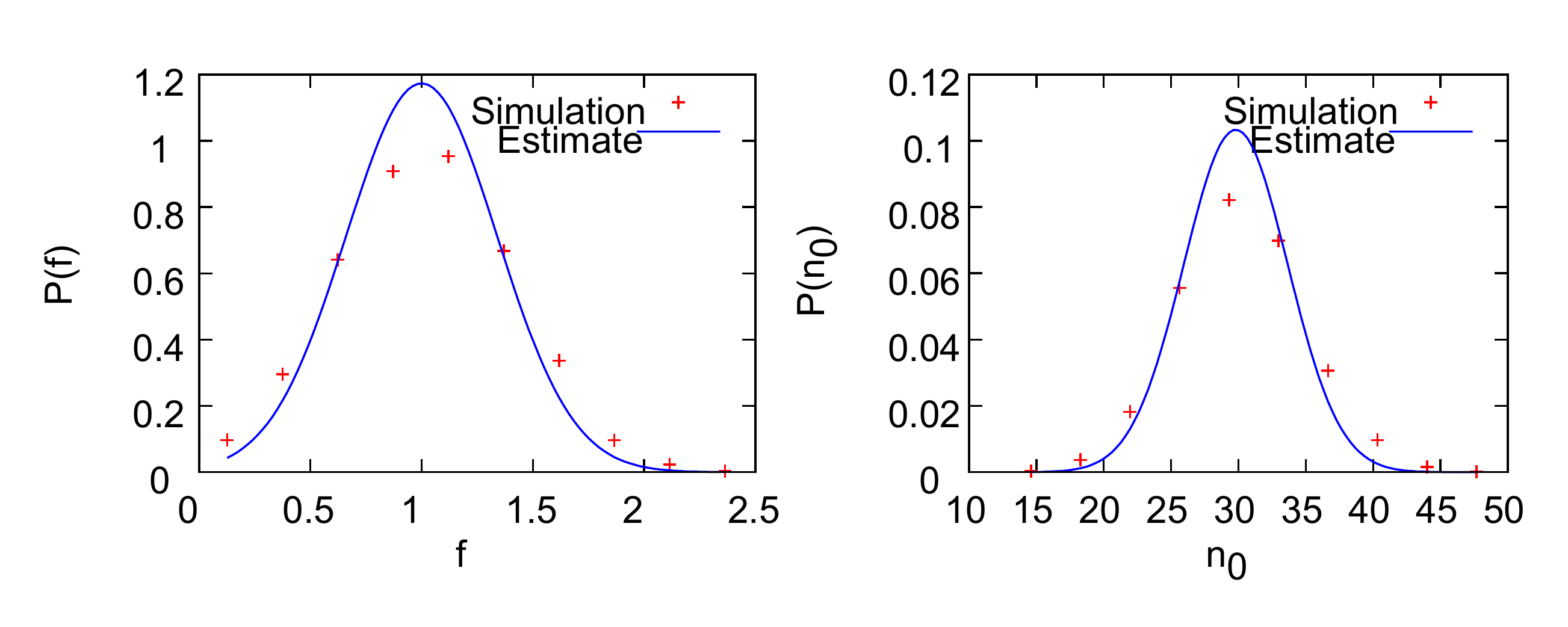}
\caption{A. Probability distribution of $n_0$ under default model parameters, compared to the estimated distribution from considering a population-average cell. B. Probability distribution of $f$ under default model parameters, compared to the distribution of the underlying AR(1) process.}
\label{fdist}
\end{figure*}

\begin{figure}
\includegraphics[width=8cm]{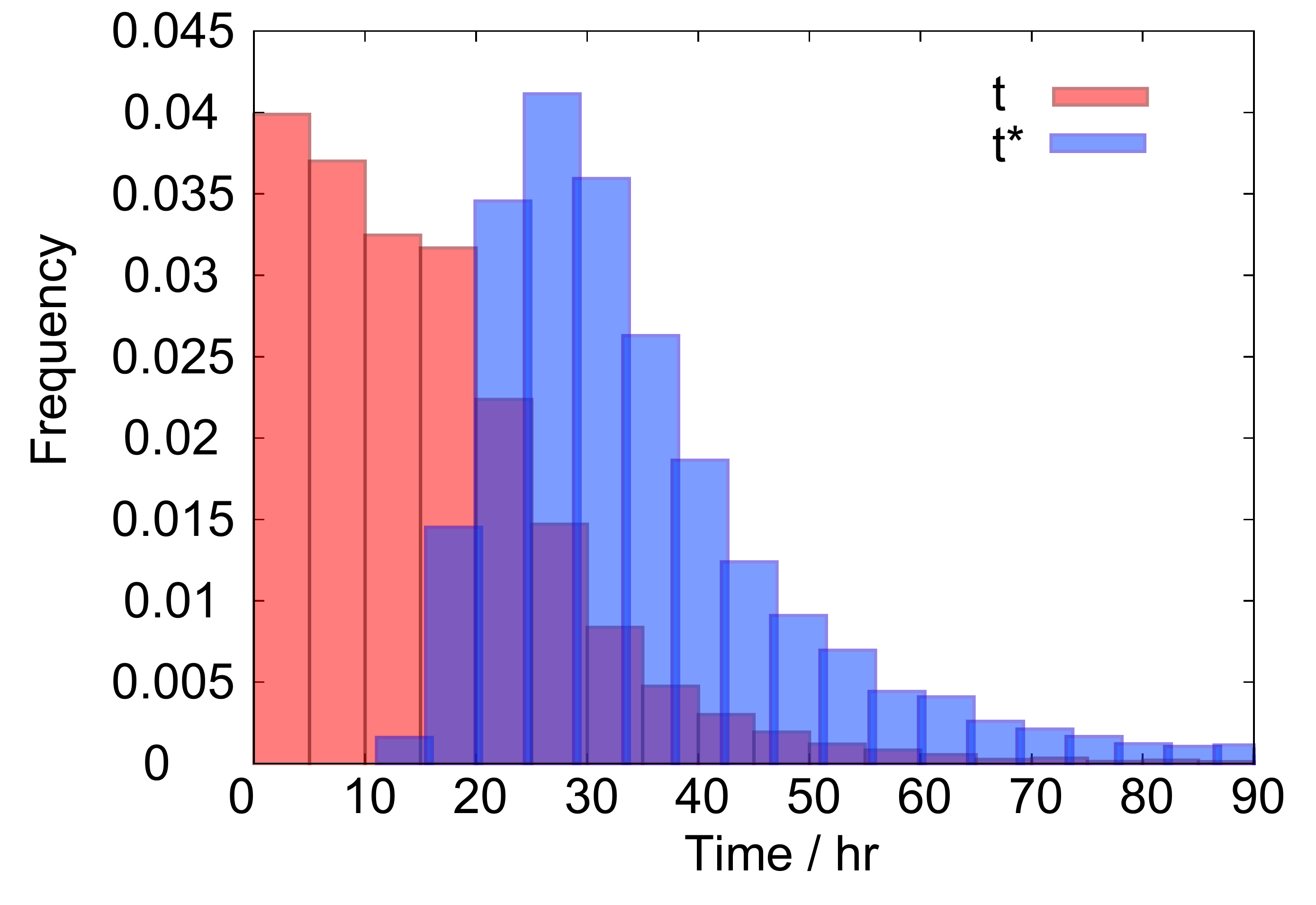} 
\caption{Probability distributions of cellular age $t$ and cell cycle length $t^*$ under default model parameters.}
\label{timedist}
\end{figure}

\subsection{Other Models}
\subsubsection{Continuous $f$}
The behaviour of mitochondrial functionality $f$ throughout and between cell cycles is the least well characterised property of our model. We use a simple picture in which $f$ stays constant throughout a cell cycle and changes stochastically at mitosis, chosen for simplicity and due to the observed slowly varying behaviour of membrane potential in the cell cycle \cite{neves2010connecting}. An alternative picture involves $f$ being allowed to vary continuously and randomly throughout the cell cycle and no discrete jump occuring at mitosis, with daughter cells straightforwardly inheriting their parent's functionality. We found that this system was difficult to approach analytically, but performed numerical experiments to explore its behaviour. Instead of parameterising the behaviour of $f$ with $f_a$ and $f_c$, describing an autoregressive process, we now allow $f$ to vary in a random walk with steps of size $\delta f$ per unit time. $\delta f$ then measures the degree of variability associated with $f$. To avoid unphysical situations, we place the constraint $0.1 \leq f < 1.9$, forcing $f$ to follow a constrained random walk.

\begin{figure}
\includegraphics[width=8cm]{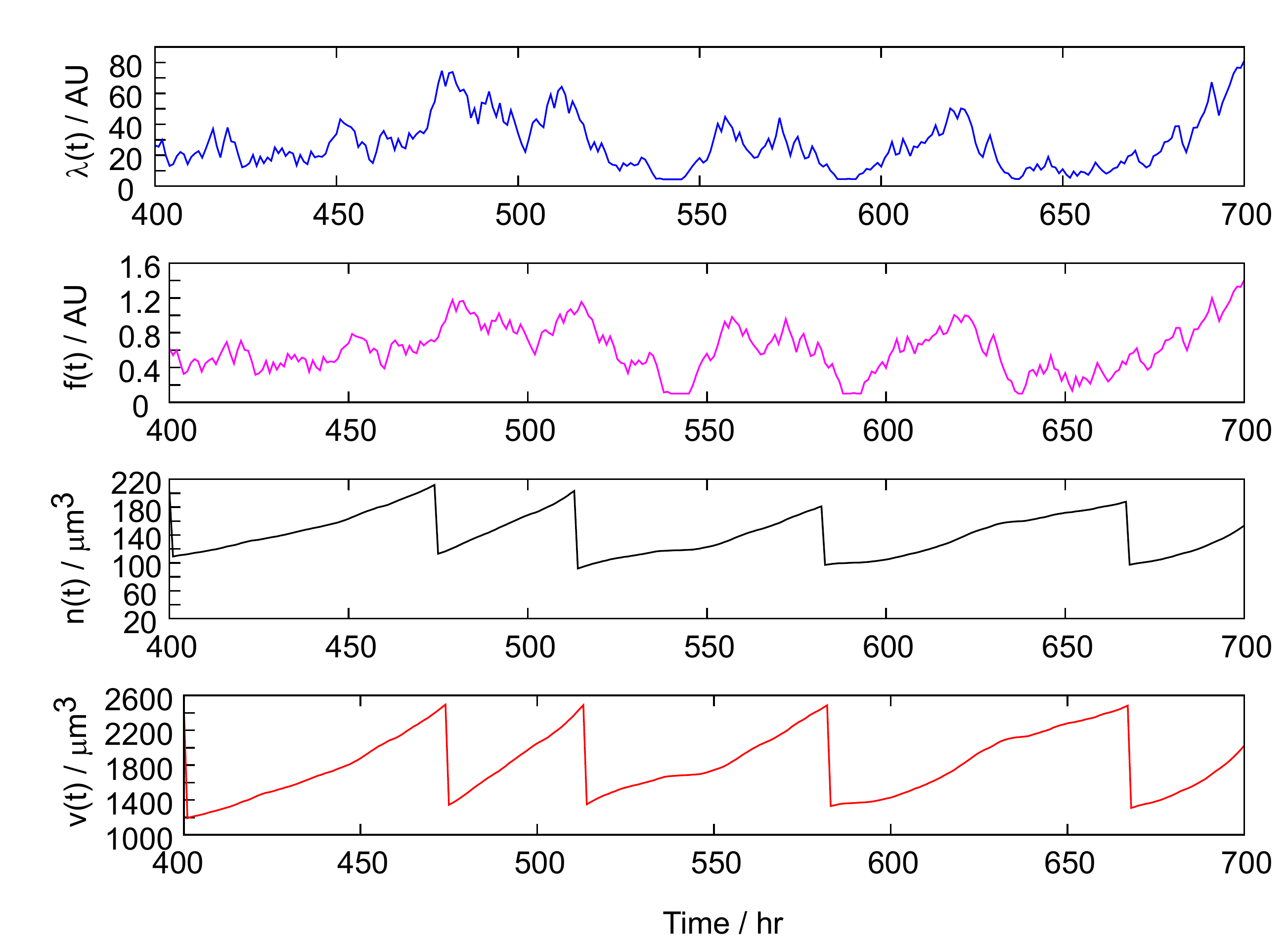}
\caption{Time series with varying $f$.}
\label{varyingfts}
\end{figure}

\begin{figure*}
\includegraphics[width=17cm]{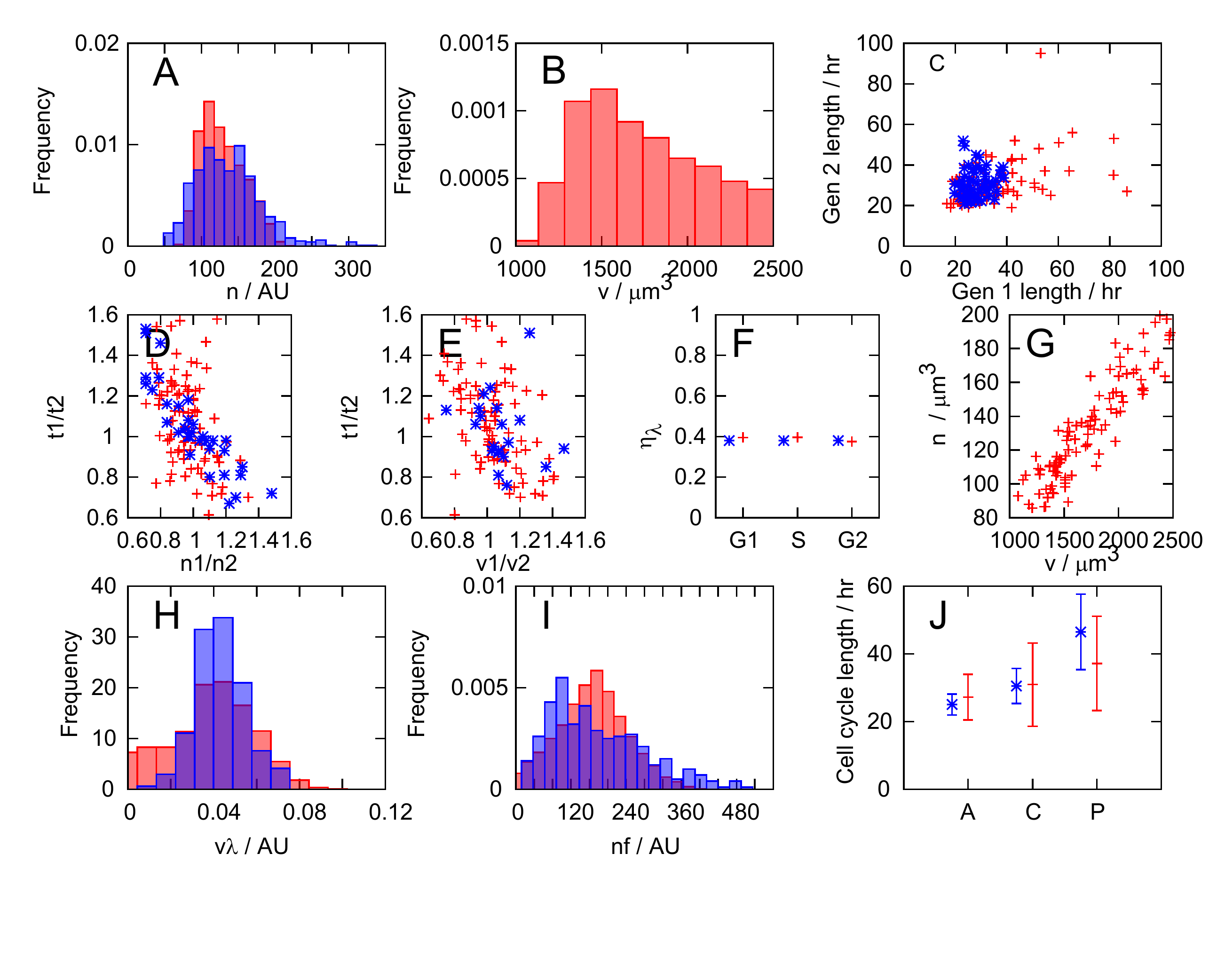}
\caption{\textbf{Behaviour of our model with continuously varying $f$.} Comparison to experimental data of the model with $f$ continuously varying within cell cycles and inherited continuously at mitosis. Plots are the same as Fig. \ref{wufig2} in the Main Text.}
\label{varyf}
\end{figure*}

Figs. \ref{varyingfts} and \ref{varyf} shows the behaviour of this model after parameterisation on the same data used to parameterise our default model. Most experimental results are captured to a similar extent than observed with our default model. These results suggest that the model's agreement with experimental data is robust with respect to the fine detail of the time evolution of $f$ within cell cycles. This robustness suggests that it is the magnitude of extrinsic variability in mitochondrial functionality that gives rise to important physiological effects, rather than the specific detail of the time evolution and inheritance of functionality. However, as different dynamic regimes may adequately represent the behaviour of mitochondrial functionality, the need for further elucidation of mitochondrial functionality in model building is emphasised.

\subsubsection{\label{atpstuff} ATP and Alternatives}
It is thought that ATP levels in the cell are subject to homeostasis. For this reason we do not include a sink term for ATP, assuming that an ATP deficit will be immediately compensated for. A model that would capture ATP usage is given by:

\begin{eqnarray}
\dot{v} & = & \alpha' \frac{a}{v} \\
\dot{n} & = & \beta' \frac{a}{v} \\
\dot{a} & = & \gamma' n - (\alpha'+\beta') \frac{a}{v}, \label{atpsink}
\end{eqnarray}

where $a$ is the cellular ATP level. Here, the second term in Eqn. \ref{atpsink} corresponds to the rate of use of ATP in increasing the cell volume and mitochondrial mass. Note that if $a = \gamma n$ our model equations are recovered. This model, with a suitable parameterisation, produces very similar results to our simpler model.

We also note that the following system explicitly incorporates ATP homeostasis, as ATP is produced up to a certain level by mitochondria and used up in the increase of cellular volume:

\begin{equation}
\dot{a} = \alpha' (a^* - a) n - \beta' a v,
\end{equation}

where $a^*$ controls the homeostatic level of ATP. This equation is solved by

\begin{equation}
a = a_0 e^{-t(\alpha' n - \beta' v)} + \frac{a^* \alpha' n}{\alpha' n + \beta' v},
\end{equation}

giving a hyperbolic form for ATP levels in terms of $n$ and $v$. This representation could easily be extended to allow consideration of the ATP:ADP ratio -- in which the rate of use of ATP gives the rate of production of ADP and vice versa.

It is also possible that ROS plays a role in modulating transcription rate and cellular growth rates. In this case, an additional term could be introduced into the model, proportional to the number of mitochondria, and possibly inversely proportional to mitochondrial functionality, and used to modulate the dynamic equations and transcription rate.

In our current model, we do not consider these complicating factors, due to our lack of experimental justification for them and also due to our desire to retain a simple, analytically tractable model. Future work, possibly motivated by the experiments we suggest in the main text, could take these more complicated factors into account.

\subsubsection{Other Deterministic Dynamic Forms}
\begin{figure*}
\includegraphics[width=17cm]{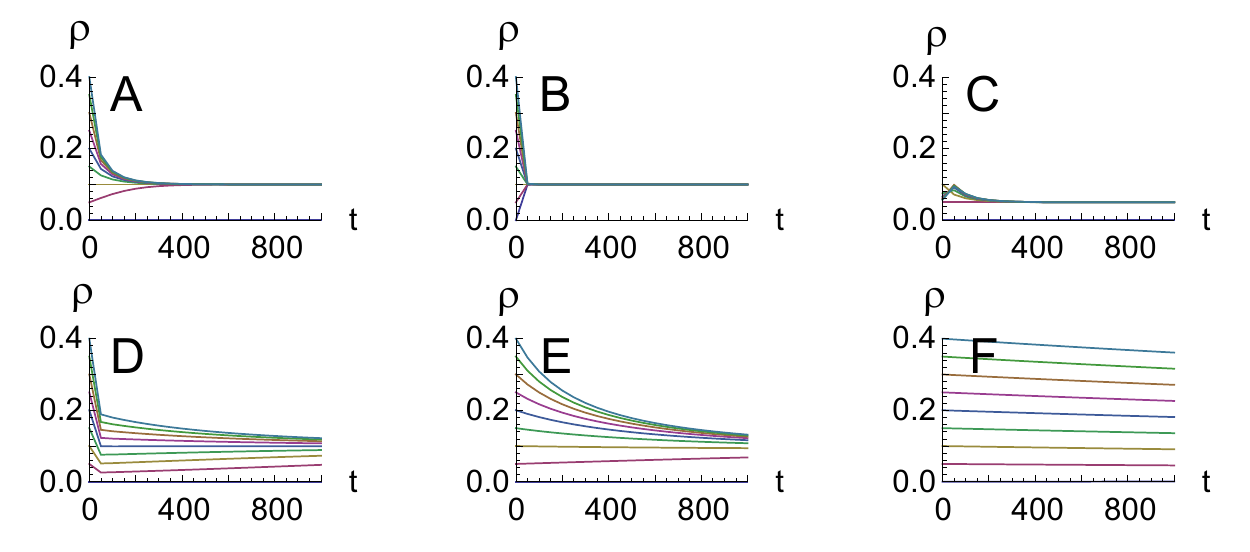}
\caption{The functional form of mitochondrial density $\rho$ with time $t$ (in arbitrary units) in other possible models for the time evolution of cellular properties. For these illustrations, the parameter set $\alpha = 0.1, \beta = 0.01, v_0 = 1000$ is used. The different lines correspond to trajectories of $\rho$ resulting from varying initial $n_0$ between 0 and 400 (a much wider range than in the default model, as different choices of functional form may lead to different absolute values for $n$ and $v$). The label of the plot gives the corresponding function in Table \ref{diffmods}.}
\label{othermods}
\end{figure*}

Our model has been chosen phenomenologically to match experimental results concerning the distribution and behaviour of mitochondria. Other models are possible that display similar behaviour. Here we mention some potential alternative models. These alternatives both incorporate mean reversion on mitochondrial density and yield sensible key results. There is a difference in mean reversion rate between these models, illustrated by dynamic plots in Fig. \ref{othermods}. 

\begin{table}
\begin{tabular}{|c|l|p{8cm}|}
\hline
Label & Equations & Description \\
\hline\hline
A & $\begin{array}{rl} \dot{v} & =  \alpha n \\ \dot{n} & =  \beta n \end{array}$ & \small Default model type \\
\hline
B & $\begin{array}{rl} \dot{v} & =  \alpha v \\ \dot{n} & =  \beta v \end{array}$ & \small Volume control on growth \\
\hline
C & $\begin{array}{rl} \dot{v} & =  \alpha n \\ \dot{n} & =  \alpha n \left( \gamma - \frac{n}{v} \right) \end{array}$ & \small Strong, explicit mean reversion of $n$ towards a density $\gamma$. \\
\hline
D & $\begin{array}{rl} \dot{v} & =  \alpha v \left( 1 - \frac{v}{v^*} \right) \\ \dot{n} & =  \alpha n \left( \gamma - \frac{n}{v} \right) \end{array}$ & \small Explicit mean reversion of $n$ towards $\gamma$, and volume growth towards a target $v^*$ \\
\hline
E & $\begin{array}{rl} \dot{v} & =  \alpha \\ \dot{n} & =  \beta n \left( \gamma - \frac{n}{v} \right) \end{array}$ & \small Linear volume growth \\
\hline
F & $\begin{array}{rl} \dot{v} & =  \alpha \\ \dot{n} & =  \beta \left( \gamma - \frac{n}{v} \right) \end{array}$ & Linear volume growth and weaker mean reversion on $n$ \\
\hline
\end{tabular}
\caption{Different possible models for time evolution of the cell exhibiting mean reversion. Each of these expressions admits an analytic solution for $n(t)$ and $v(t)$. Other combinations (for example, weak mean reversion on $n$ and exponential volume growth) are harder to solve and are often unstable for high or low initial $n/v$.}
\label{diffmods}
\end{table}

Our model was chosen for its analytic tractability and its ability to reproduce experimental phenomena: these more complex models rapidly become analytically intractable, which hinders a deeper understanding of their behaviour.

We note in addition that the $[ATP]$ term that appears in our dynamic equations may be replaced by a transcription or translation rate term, explicitly including the sigmoidal (or hyperbolic) response of these cellular growth rates in the dynamic equations. In our bare model, we use the simpler $[ATP]$ term both to maintain analytic tractability and because the appropriate form for a response curve is difficult to estimate (for example, cellular growth may be a result of a combination of transcription and translation, and may include other additional terms). This approach essentially involves using a linear approximation for growth rate: linear approximations for transcription rate are employed elsewhere in this study and yield very similar results to the numerical behaviour of the corresponding nonlinear case.

It is possible to construct models for mitochondrial stochasticity that do not involve mean-reverting behaviour of mitochondrial density. As a simple example, if cellular dynamics are such that mitochondrial mass and cellular volume evolve independently:

\begin{eqnarray}
\dot{v} & = & \alpha f v; \\
\dot{n} & = & \beta f n,
\end{eqnarray}

with mitosis involving a binomial segregation of volume and mitochondrial mass as before, the population variance in mitochondrial density will generally increase in an unbounded manner, as the probability of observing higher and higher mitochondrial densities increases with time. However, if cutoffs are placed on mitochondrial density, so that cells are removed from the population if their density does not obey

\begin{equation}
\rho_{-} < \frac{n}{v} < \rho_{+},
\end{equation}
  
the population distribution of mitochondrial density can achieve stationarity. This approach results in a much weaker correlation between mitochondrial mass $n$ and cellular volume $v$. This model can be parameterised (an example parameter set is $\rho_{-} = 10, \rho_{+} = 300, \alpha = \beta = 0.1, \gamma = 3.6$, other parameters the same as in the default model) to yield similar results to the default model for transcriptional noise.


This rather constructed model is considered due to the discrepancy between flow cytometry data in das Neves \emph{et al.} \cite{neves2010connecting}, showing an absence of correlation between mitochondrial volume and cellular volume. Obtaining analytic results from this model is difficult, both in terms of approximation for the moments of probability distributions and the time evolution of mRNA. 


\subsection{Potential Experiments for Refinement}
As mentioned in the Main Text, our model was constructed from a phenomenological philosophy, with the intention of using experimental results to construct a plausible coarse-grained explanation for the influence of mitochondrial variability on extrinsic noise in general and transcription rate in particular. Our goal was to introduce a simplified but consistent mathematical summary of the data and to use this to motivate further experiments. We suggest a set of experiments in Table II that would support or contribute to further development of this model.

\begin{table*}
\small{
\begin{tabular}{|c|p{6cm}|p{8cm}|}
\hline
& Observable & Prediction \\
\hline\hline
1 & Mitochondrial functionality (measured via membrane potential) before and after mitosis. & Parent membrane potential is weakly retained but stochastically altered to give daughter membrane potential. Note that our current model treats this process crudely and as such many other dynamics may be possible. \\
2 & Protein expression levels as a function of $[ATP]$. & Expression levels are higher in cells with higher $[ATP]$. \\
3 & Time series of mitochondrial mass and density through the cell cycle. & Exponential cell growth, with mitochondrial density tending towards an average value. The time series of these dynamics could be used to distinguish between different models (see `Other Models'). \\
4 & Behaviour of $[ATP]$ with time. & Slowly varying over the cell cycle, mean-reverting towards an average value. \\
5 & Determinant factors of $[ATP]$. & Mitochondrial mass and membrane potential, and cell volume. Levels of ROS may also be of importance. \\
6 & Relative contribution of mitochondrial mass and functional variability to transcriptional variability (measured via bromouridine uptake). & Strong dependence of transcriptional noise on both mitochondrial inheritance and functionality variability. \\
7 & Noise in protein expression levels. & Dependent on mitochondrial variability, and lower for cells with mitochondrial low mass and functionality. \\
8 & ROS levels (measured via, for example, MitoSox). & Possibly correlated inversely with a measure of mitochondrial functionality, and possibly higher with low mitochondrial mass and functionality, as weaker/sparser mitochondria struggle to match energy demands.  \\
\hline
\end{tabular}
}
\label{expts1}
\caption{Potential experiments that may clarify aspects of our model, roughly ranked in order of importance for supporting or suggesting area of refinement for our model. }
\end{table*}

\subsection{mRNA \& Protein Levels}
The master equation for the probability distribution of mRNA levels can be written down as:

\begin{equation}
\frac{\partial P_m}{\partial t} = \lambda(t) P_{m-1} + \xi (m+1) P_{m+1} - (\lambda(t) + \xi m) P_m,
\end{equation}

where $\lambda(t)$ is the time-dependent birth rate of mRNA molecules, $\xi$ is the rate of removal of mRNA, and $P_m(t)$ is the probability of observing the system with $m$ mRNA molecules at time $t$. We choose



\begin{equation}
\lambda(t) = (c + b t),
\end{equation}

the approximation from `Parameterisation of $\lambda(t)$', to model transcription rate with time.

Using the generating function $G(z) = \sum_m z^m P_m$ gives:

\begin{equation}
\frac{\partial G}{\partial t} = \lambda(t) (z - 1) G - \xi (z-1) \frac{\partial G}{\partial z}.
\label{geqn}
\end{equation}

We can solve Eqn. \ref{geqn} with the method of characteristics. This process allows us to convert the PDE into a set of ODEs along a characteristic curve of the function. We wish to recast Eqn. \ref{geqn} into the following form, where a characteristic curve is parameterised by the new variable $s$:

\begin{equation}
\frac{d}{ds} G(z(s), t(s)) = F(G, z(s), t(s)).
\end{equation}

Using the chain rule, we can write:

\begin{eqnarray}
\frac{d}{ds} G(z(s), t(s)) & = & \frac{\partial G}{\partial z} \frac{dz}{ds} + \frac{\partial G}{\partial t} \frac{dt}{ds} \\
& = & \frac{1}{z-1} \frac{\partial G}{\partial t} + \xi \frac{\partial G}{\partial z} = (c + bt) G,
\end{eqnarray}

where the last line is just a rearrangement of the original PDE. By comparing coefficients, we obtain the characteristic equations for the system:

\begin{eqnarray}
\frac{dz}{ds} & = & \xi \, \rightarrow \, z = \xi s + z_0 \\
\frac{dt}{ds} & = & \frac{1}{z-1} = \frac{1}{\xi s + z_0 - 1} \, \rightarrow \, t = \frac{1}{\xi} \ln (\xi s + z_0 - 1) + t_0, \\
\frac{dG}{ds} & = & (c + bt) G = \left( c + b \left( \frac{1}{\xi} \ln (\xi s + z_0 - 1) + t_0 \right) \right) G.
\end{eqnarray}

These equations may be interpreted as describing the evolution of the arguments of $G$, and the subsequent evolution of the value of $G$, as we progress along a curve parameterised by $s$. As absolute values of $s$ are not important, affecting only the parameterisation of progress along a characteristic curve, we set $z_0 = 0$ without loss of generality. We then have $s = \frac{z}{\xi}$ and $t_0 = t - \frac{1}{\xi} \ln (z-1)$. The final ODE can then be solved by separation of variables:

\begin{eqnarray}
\int \frac{1}{G} dG & = & \int \left( c + b \left( \frac{1}{r} \ln (\xi s + z_0 - 1) + c_2 \right) \right) ds \\
& = & s(c+b t_0) - \frac{s b}{\xi} + \frac{s b}{\xi} \ln (\xi s -1) - \frac{b}{\xi^2} \ln (\xi s - 1) \\
& = & \frac{z}{\xi} (c + bt) - \frac{bz}{\xi^2} - \frac{b}{\xi^2} \ln (z-1),
\end{eqnarray}

We then have

\begin{equation}
G = C \exp \left( \frac{z}{\xi}(c+bt) - \frac{bz}{\xi^2} \right) (z-1)^{\frac{-b}{\xi^2}},
\end{equation}

where the arbitrary function $C = C(t - \frac{1}{\xi} \ln (z-1)) = C(t_0)$, as this quantity is independent of $s$.

If the initial copy number of mRNA molecules is $m_0$, we have the initial condition $G(z, 0) = \sum_m z^m P_m(0) = \sum_m z^m \delta_{m\,m_0} = z^{m_0}$. Noting that at $t = 0$, $e^{-t_0 \xi} = z-1$, we find that if we employ the choice

\begin{equation}
C(t_0) = \exp \left( (e^{-t_0 \xi} + 1) \left(\frac{b}{\xi^2} - \frac{c}{\xi} \right) \right) (e^{-t_0 \xi}+1)^{m_0} \exp \left( \frac{-b}{\xi^2} t_0 \xi \right),
\end{equation}

we recover the required initial condition:

\begin{eqnarray}
G(z,t_0) & = & \exp \left( \frac{z}{\xi}(c+bt) - \frac{bz}{\xi^2} \right) (z-1)^{\frac{-b}{\xi^2}} \exp \left( (e^{-t_0 \xi} + 1) \left(\frac{b}{\xi^2} - \frac{c}{\xi} \right) \right) e^{-t_0 \xi}+1)^{m_0} \exp \left( \frac{-b}{\xi^2} t_0 \xi \right) \label{fullsoln} \\
G(z, t_0 = 0) & = & \exp \left( \frac{z c}{\xi} - \frac{b z}{\xi^2} \right) (z-1)^{\frac{-b}{\xi^2}} \exp \left( z \left(\frac{b}{\xi^2} - \frac{c}{\xi} \right) \right) z^{m_0} (z-1)^{\frac{b}{\xi^2}} \\
& = & z^{m_0}.
\end{eqnarray}

The general solution is then given by Eqn. \ref{fullsoln}, which, using $t_0 = t - \frac{1}{\xi} \ln (z-1)$, can be written:

\begin{equation}
G(z,t) = e^{a_1 z + a_2} (z+a_3)^{m_0},
\end{equation}

with

\begin{eqnarray}
a_1 & = & \frac{1}{\xi} \left(c + bt - c e^{-\xi t} + \frac{b}{\xi}(e^{-\xi t} - 1) \right) \\
a_2 & = & -m_0 t \xi - a_1 \\
a_3 & = & e^{\xi t} - 1
\end{eqnarray}

We can recover the mean and variance of the distribution $P_m(t)$ by using $\mu_m = \left. \frac{\partial G}{\partial z} \right|_{z = 1}$ and $\sigma^2_m = \left. \frac{\partial^2 G}{\partial z^2}\right|_{z = 1} + \left. \frac{\partial G}{\partial z}\right|_{z = 1} - \left(\left. \frac{\partial G}{\partial z}\right|_{z = 1} \right)^2$:

\begin{eqnarray}
\mu_m & = & (a_3 + 1)^{m_0 - 1} e^{a_1 + a_2} (a_1 + a_1 a_3 + m_0) \\
\sigma^2_m & = & (a_3 + 1)^{m_0} e^{a_1 + a_2} \left( a_1 + a_1^2 + \frac{m_0}{a_3 + 1} \left( 1 + 2 a_1 + \frac{m_0 - 1}{a_3 + 1} \right) - (a_3 + 1)^{m_0 + 2} e^{a_1 + a_2} (a_1 + a_1 a_3 + m_0)^2 \right)
\end{eqnarray}

We can also recover the full probability distribution of observing $m$ mRNAs using $P_m(t) = \left. \frac{1}{m!} \frac{\partial^m G}{\partial z^m} \right|_{z = 0}$. Using Leibniz's rule allows us to write:

\begin{eqnarray}
P_m(t) & = & \left. \frac{1}{m!} \frac{\partial^m G}{\partial z^m} \right|_{z = 0} \\
& = & \left. \frac{1}{m!} \sum_{i = 0}^m \binom{m}{i} \frac{\partial^i}{\partial z^i} e^{a_1 z + a_2} \frac{\partial^{m-i}}{\partial z^{m-i}} (z + a_3)^{m_0} \right|_{z = 0} \\
& = & \left. \sum_{i = 0}^m \frac{1}{i! (m-i)!} a_1^i e^{a_1 z + a_2} \frac{m_0!}{(m_0 - m + i)!} (z + a_3)^{m_0 - m + i} \right|_{z = 0} \\
& = & \sum_{i = 0}^m \frac{m_0! a_1^i e^{a_2} a_3^{m_0 - m + i}}{(m_0 - m + i)! i! (m-i)!}.
\end{eqnarray}

This can be alternatively written as

\begin{equation}
P_m(t) = a_3^{m_0-m} e^{a_2} \frac{m_0!}{m! (m_0 - m)!} {}_1F_1(-m; m_0 - m + 1; -a_1 a_3)
\end{equation}

where ${}_1 F_1$ is the Kummer confluent hypergeometric function.

The corresponding master equation including proteins is:

\begin{eqnarray}
\frac{\partial P_{m\,n}}{\partial t} & = & m \lambda_n(t) P_{m\,n-1} + \xi_n (n+1) P_{m\,n+1} - \xi_n n P_{m\,n} - m \lambda_n(t) P_{m\,n} \nonumber \\
&& + \lambda_m(t) P_{m-1\,n} +  \xi_m (m+1) P_{m+1\,n} - \xi_m m P_{m\,n} - \lambda_m(t) P_{m\,n},
\end{eqnarray}

for which we have not obtained a general analytic solution. Instead, we simulate this system with a stochastic simulation algorithm. To simulate these systems, we use a parameter set employed by Raj \emph{et al.} \cite{raj2006stochastic}: $\langle \lambda_m \rangle = 0.06\,s^{-1}, \langle \lambda_n \rangle = 0.007\,s^{-1}, \xi_m = 7 \times 10^{-5}\,s^{-1}, \xi_n = 1.1 \times 10^{-5}\,s^{-1}$. The death rates are simply inserted into the simulation, and the mean birth rates are used to tune the time-varying birth rate for the simulation. For example, the normalisation of $\lambda_m(t)$ was chosen so that the population average value of $[ATP]$ corresponded to $\langle \lambda_m \rangle$.

As the process of translation is believed to be $[ATP]$-dependent but not dependent on chromatin remodelling, the decondensed version of Eqn. \ref{nuatp} (using the hyperbolic $s^d_i$ coefficients rather than the sigmoidal $s_i$ coefficients) was used for $\lambda_n$. Again, the mean birth rate was used to normalise this function as above. Overall, we then have constant rates for mRNA death (faster) and protein death (slower) and time-varying birth rates for mRNA (including chromatin remodelling) and proteins (not including chromatin remodelling).

As mentioned in the Main Text, we found this parameterisation to yield large copy numbers of proteins, which led to very low values for intrinsic noise. We explored this effect by increasing the degradation rates from Raj \emph{et al.}'s default values, to $\xi_m = 7 \times 10^{-3}\,s^{-1}, \xi_n = 1.1 \times 10^{-3}$. In addition, we investigated the case where variability in $[ATP]$ only affected transcription rate, with translation rate taking the constant value $\lambda_n = 0.007\,s^{-1}$. As shown in the Main Text, those simulations with higher degradation rates showed a more significant intrinsic noise contribution, and the trends in these dual reporter simulations changed little with a constant translation rate. This consistency means that though our crude model of the dependence of translation rate on $[ATP]$ could be challenged, even if translation is ATP dependent we still see pronounced effects on protein levels through ATP-dependent variability in transcription rate.





\bibliographystyle{unsrt}
\bibliography{refs1}

\begin{thebibliography}{10}

\bibitem{mcadams1997stochastic}
H.~H. McAdams and A.~Arkin.
\newblock {Stochastic mechanisms in gene expression}.
\newblock {\em Proc. Natl. Acad. Sci. USA}, 94(3):814--819, 1997.

\bibitem{altschuler2010cellular}
S.J. Altschuler and L.F. Wu.
\newblock {Cellular Heterogeneity: Do Differences Make a Difference?}
\newblock {\em Cell}, 141(4):559--563, 2010.

\bibitem{elowitz2002stochastic}
M.~B. Elowitz, A.~J. Levine, E.~D. Siggia, and P.~S. Swain.
\newblock {Stochastic gene expression in a single cell}.
\newblock {\em Science}, 297(5584):1183--1186, 2002.

\bibitem{kaern2005stochasticity}
M.~K{\ae}rn, T.~C. Elston, W.~J. Blake, and J.~J. Collins.
\newblock {Stochasticity in gene expression: from theories to phenotypes}.
\newblock {\em Nat. Rev. Genet.}, 6(6):451--464, 2005.

\bibitem{raj2008nature}
A.~Raj and A.~van Oudenaarden.
\newblock {Nature, nurture, or chance: stochastic gene expression and its
  consequences}.
\newblock {\em Cell}, 135(2):216--226, 2008.

\bibitem{chang2008transcriptome}
H.H. Chang, M.~Hemberg, M.~Barahona, D.E. Ingber, and S.~Huang.
\newblock {Transcriptome-wide noise controls lineage choice in mammalian
  progenitor cells}.
\newblock {\em Nature}, 453(7194):544--547, 2008.

\bibitem{fraser2009chance}
D.~Fraser and M.~K{\ae}rn.
\newblock {A chance at survival: gene expression noise and phenotypic
  diversification strategies}.
\newblock {\em Mol. Microbiol.}, 71(6):1333--1340, 2009.

\bibitem{kussell2005bacterial}
E.~Kussell, R.~Kishony, N.Q. Balaban, and S.~Leibler.
\newblock {Bacterial persistence: a model of survival in changing
  environments}.
\newblock {\em Genetics}, 169(4):1807, 2005.

\bibitem{brock2009non}
A.~Brock, H.~Chang, and S.~Huang.
\newblock {Non-genetic heterogeneity -- a mutation-independent driving force
  for the somatic evolution of tumours}.
\newblock {\em Nat. Rev. Genet.}, 10(5):336--342, 2009.

\bibitem{bastiaens2009systems}
P.~Bastiaens.
\newblock {Systems biology: when it is time to die}.
\newblock {\em Nature}, 459(7245):334--335, 2009.

\bibitem{spencer2009non}
S.L. Spencer, S.~Gaudet, J.G. Albeck, J.M. Burke, and P.K. Sorger.
\newblock {Non-genetic origins of cell-to-cell variability in TRAIL-induced
  apoptosis}.
\newblock {\em Nature}, 459(7245):428--432, 2009.

\bibitem{swain2002intrinsic}
P.S. Swain, M.B. Elowitz, and E.D. Siggia.
\newblock {Intrinsic and extrinsic contributions to stochasticity in gene
  expression}.
\newblock {\em Proc. Natl. Acad. Sci. USA}, 99(20):12795, 2002.

\bibitem{hilfinger2011separating}
A.~Hilfinger and J.~Paulsson.
\newblock Separating intrinsic from extrinsic fluctuations in dynamic
  biological systems.
\newblock {\em Proc. Natl. Acad. Sci. USA}, 108(29):12167, 2011.

\bibitem{raser2004control}
J.~M. Raser and E.~K. O'Shea.
\newblock {Control of stochasticity in eukaryotic gene expression}.
\newblock {\em Science}, 304(5678):1811--1814, 2004.

\bibitem{newman2006single}
J.~R.~S. Newman, S.~Ghaemmaghami, J.~Ihmels, D.~K. Breslow, M.~Noble, J.~L.
  DeRisi, and J.~S. Weissman.
\newblock {Single-cell proteomic analysis of S. cerevisiae reveals the
  architecture of biological noise}.
\newblock {\em Nature}, 441(7095):840--846, 2006.

\bibitem{raj2006stochastic}
A.~Raj, C.~S. Peskin, D.~Tranchina, D.~Y. Vargas, and S.~Tyagi.
\newblock {Stochastic mRNA synthesis in mammalian cells}.
\newblock {\em PLoS Biol.}, 4(10):e309, 2006.

\bibitem{paulsson2005models}
J.~Paulsson.
\newblock {Models of stochastic gene expression}.
\newblock {\em Phys. Life Rev.}, 2(2):157--175, 2005.

\bibitem{paulsson2004summing}
J.~Paulsson.
\newblock {Summing up the noise in gene networks}.
\newblock {\em Nature}, 427(6973):415--418, 2004.

\bibitem{volfson2005origins}
D.~Volfson, J.~Marciniak, W.~J. Blake, N.~Ostroff, L.~S. Tsimring, and
  J.~Hasty.
\newblock {Origins of extrinsic variability in eukaryotic gene expression}.
\newblock {\em Nature}, 439(7078):861--864, 2005.

\bibitem{bruggeman2009noise}
F.J. Bruggeman, N.~Bl{\"u}thgen, and H.V. Westerhoff.
\newblock {Noise management by molecular networks}.
\newblock {\em PLoS Comp. Biol.}, 5(9):1183--1186, 2009.

\bibitem{rausenberger2008quantifying}
J.~Rausenberger and M.~Kollmann.
\newblock {Quantifying origins of cell-to-cell variations in gene expression}.
\newblock {\em Biophys. J.}, 95(10):4523--4528, 2008.

\bibitem{blake2003noise}
W.~J. Blake, M.~K{\ae}rn, C.~R. Cantor, and J.~J. Collins.
\newblock {Noise in eukaryotic gene expression}.
\newblock {\em Nature}, 422(6932):633--637, 2003.

\bibitem{dobrzynski2009elongation}
M.~Dobrzy{\'n}ski and F.~J. Bruggeman.
\newblock {Elongation dynamics shape bursty transcription and translation}.
\newblock {\em Proc. Natl. Acad. Sci. USA}, 106(8):2583--2588, 2009.

\bibitem{thattai2001intrinsic}
M.~Thattai and A.~Van~Oudenaarden.
\newblock {Intrinsic noise in gene regulatory networks}.
\newblock {\em Proc. Natl. Acad. Sci. USA}, 98(15):8614, 2001.

\bibitem{sigal2006variability}
A.~Sigal, R.~Milo, A.~Cohen, N.~Geva-Zatorsky, Y.~Klein, Y.~Liron,
  N.~Rosenfeld, T.~Danon, N.~Perzov, and U.~Alon.
\newblock {Variability and memory of protein levels in human cells}.
\newblock {\em Nature}, 444(7119):643--646, 2006.

\bibitem{sigal2006dynamic}
A.~Sigal, R.~Milo, A.~Cohen, N.~Geva-Zatorsky, Y.~Klein, I.~Alaluf,
  N.~Swerdlin, N.~Perzov, T.~Danon, Y.~Liron, et~al.
\newblock {Dynamic proteomics in individual human cells uncovers widespread
  cell-cycle dependence of nuclear proteins}.
\newblock {\em Nature Methods}, 3(7):525--531, 2006.

\bibitem{bareven2006noise}
A.~Bar-Even, J.~Paulsson, N.~Maheshri, M.~Carmi, E.~O'Shea, Y.~Pilpel, and
  N.~Barkai.
\newblock {Noise in protein expression scales with natural protein abundance}.
\newblock {\em Nature Genetics}, 38(6):636--643, 2006.

\bibitem{kaufmann2007stochastic}
B.B. Kaufmann and A.~van Oudenaarden.
\newblock {Stochastic gene expression: from single molecules to the proteome}.
\newblock {\em Curr. Opin. Gen. Dev.}, 17(2):107--112, 2007.

\bibitem{huh2010non}
D.~Huh and J.~Paulsson.
\newblock {Non-genetic heterogeneity from stochastic partitioning at cell
  division}.
\newblock {\em Nat. Genetics}, 43(2):95--102, 2010.

\bibitem{huh2011random}
D.~Huh and J.~Paulsson.
\newblock Random partitioning of molecules at cell division.
\newblock {\em Proc. Natl. Acad. Sci. USA}, 108(36):15004, 2011.

\bibitem{neves2010connecting}
R.~P. das Neves, N.~S. Jones, L.~Andreu, R.~Gupta, T.~Enver, and F.~J. Iborra.
\newblock {Connecting Variability in Global Transcription Rate to Mitochondrial
  Variability}.
\newblock {\em PLoS Biol.}, 8(12):451--464, 2010.

\bibitem{mcbride2006mitochondria}
H.M. McBride, M.~Neuspiel, and S.~Wasiak.
\newblock {Mitochondria: more than just a powerhouse}.
\newblock {\em Curr. Biol.}, 16(14):R551--R560, 2006.

\bibitem{chan2006mitochondria}
D.C. Chan.
\newblock {Mitochondria: dynamic organelles in disease, aging, and
  development}.
\newblock {\em Cell}, 125(7):1241--1252, 2006.

\bibitem{twig2008mitochondrial}
G.~Twig, B.~Hyde, and O.S. Shirihai.
\newblock {Mitochondrial fusion, fission and autophagy as a quality control
  axis: the bioenergetic view}.
\newblock {\em BBA --- Bioenergetics}, 1777(9):1092--1097, 2008.

\bibitem{collins2002mitochondria}
T.~J. Collins, M.~J. Berridge, P.~Lipp, and M.~D. Bootman.
\newblock {Mitochondria are morphologically and functionally heterogeneous
  within cells}.
\newblock {\em EMBO J.}, 21(7):1616--1627, 2002.

\bibitem{buckman2001spontaneous}
J.~F. Buckman and I.~J. Reynolds.
\newblock {Spontaneous changes in mitochondrial membrane potential in cultured
  neurons}.
\newblock {\em J. Neurosci.}, 21(14):5054--5065, 2001.

\bibitem{oreilly2003quantitative}
C.~M. O'Reilly, K.~E. Fogarty, R.~M. Drummond, R.~A. Tuft, and J.~V. Walsh.
\newblock {Quantitative analysis of spontaneous mitochondrial depolarizations}.
\newblock {\em Biophys. J.}, 85(5):3350--3357, 2003.

\bibitem{schieke2008mitochondrial}
S.~M. Schieke, M.~Ma, L.~Cao, J.~P. McCoy, C.~Liu, N.~F. Hensel, A.~J. Barrett,
  M.~Boehm, and T.~Finkel.
\newblock {Mitochondrial metabolism modulates differentiation and teratoma
  formation capacity in mouse embryonic stem cells}.
\newblock {\em J. Biol. Chem.}, 283(42):28506--28512, 2008.

\bibitem{mitra2009hyperfused}
K.~Mitra, C.~Wunder, B.~Roysam, G.~Lin, and J.~Lippincott-Schwartz.
\newblock {A hyperfused mitochondrial state achieved at G1--S regulates cyclin
  E buildup and entry into S phase}.
\newblock {\em Proc. Natl. Acad. Sci. USA}, 106(29):11960--11965, 2009.

\bibitem{mandal2005mitochondrial}
S.~Mandal, P.~Guptan, E.~Owusu-Ansah, and U.~Banerjee.
\newblock {Mitochondrial regulation of cell cycle progression during
  development as revealed by the tenured mutation in Drosophila}.
\newblock {\em Dev. Cell}, 9(6):843--854, 2005.

\bibitem{owusu2008distinct}
E.~Owusu-Ansah, A.~Yavari, S.~Mandal, and U.~Banerjee.
\newblock {Distinct mitochondrial retrograde signals control the G1-S cell
  cycle checkpoint}.
\newblock {\em Nat. Genet.}, 40(3):356--361, 2008.

\bibitem{kuznetsov2006mitochondrial}
A.V. Kuznetsov, J.~Troppmair, R.~Sucher, M.~Hermann, V.~Saks, and
  R.~Margreiter.
\newblock {Mitochondrial subpopulations and heterogeneity revealed by confocal
  imaging: Possible physiological role?}
\newblock {\em BBA Bioenergetics}, 1757(5-6):686--691, 2006.

\bibitem{cossarizza1996functional}
A.~Cossarizza, D.~Ceccarelli, and A.~Masini.
\newblock {Functional heterogeneity of an isolated mitochondrial population
  revealed by cytofluorometric analysis at the single organelle level}.
\newblock {\em Exp. Cell Res.}, 222(1):84--94, 1996.

\bibitem{mouli2009frequency}
P.K. Mouli, G.~Twig, and O.S. Shirihai.
\newblock {Frequency and selectivity of mitochondrial fusion are key to its
  quality maintenance function}.
\newblock {\em Biophys. J.}, 96(9):3509--3518, 2009.

\bibitem{shahrezaei2008colored}
V.~Shahrezaei, J.~F. Ollivier, and P.~S. Swain.
\newblock {Colored extrinsic fluctuations and stochastic gene expression}.
\newblock {\em Molecular Systems Biology}, 4(196):1--9, 2008.

\bibitem{jansen1995monte}
A.~P.~J. Jansen.
\newblock Monte carlo simulations of chemical reactions on a surface with
  time-dependent reaction-rate constants.
\newblock {\em Comp. Phys. Comm.}, 86(1):1, 1995.

\bibitem{haseltine2002approximate}
E.L. Haseltine and J.B. Rawlings.
\newblock Approximate simulation of coupled fast and slow reactions for
  stochastic chemical kinetics.
\newblock {\em J. Chem. Phys.}, 117(15):6959, 2002.

\bibitem{lane2006mitochondrial}
N.~Lane.
\newblock {Mitochondrial disease: powerhouse of disease}.
\newblock {\em Nature}, 440(7084):600--602, 2006.

\bibitem{scott2010mitochondrial}
I.~Scott and R.J. Youle.
\newblock Mitochondrial fission and fusion.
\newblock {\em Essays Biochem}, 47:85--98, 2010.

\bibitem{wilson1916distribution}
E.B. Wilson.
\newblock The distribution of the chondriosomes to the spermatozoa in
  scorpions.
\newblock {\em Proc. Natl. Acad. Sci. USA}, 2(6):321, 1916.

\bibitem{wilson1931distribution}
E.B. Wilson.
\newblock The distribution of sperm-forming materials in scorpions.
\newblock {\em J. Morphol.}, 52(2):429--483, 1931.

\bibitem{tzur2009cell}
A.~Tzur, R.~Kafri, V.~S. LeBleu, G.~Lahav, and M.~W. Kirschner.
\newblock {Cell growth and size homeostasis in proliferating animal cells}.
\newblock {\em Science}, 325(5937):167--171, 2009.

\bibitem{posakony1977mitochondrial}
J.W. Posakony, J.M. England, and G.~Attardi.
\newblock {Mitochondrial growth and division during the cell cycle in HeLa
  cells}.
\newblock {\em J. Cell Biol.}, 74(2):468--491, 1977.

\bibitem{veltri1990distinct}
K.L. Veltri, M.~Espiritu, and G.~Singh.
\newblock {Distinct genomic copy number in mitochondria of different mammalian
  organs}.
\newblock {\em J. Cell. Physio.}, 143(1):160--164, 1990.

\bibitem{herbener1976morphometric}
G.H. Herbener.
\newblock {A morphometric study of age-dependent changes in mitochondrial
  populations of mouse liver and heart}.
\newblock {\em J. Geront.}, 31(1):8, 1976.

\bibitem{mathieu1981design}
O.~Mathieu, R.~Krauer, H.~Hoppeler, P.~Gehr, S.L. Lindstedt, R.M.N. Alexander,
  C.R. Taylor, and E.R. Weibel.
\newblock {Design of the mammalian respiratory system. VII. Scaling
  mitochondrial volume in skeletal muscle to body mass}.
\newblock {\em Respir. Physio.}, 44(1):113--128, 1981.

\bibitem{suarez1991mitochondrial}
RK~Suarez, JR~Lighton, GS~Brown, and O.~Mathieu-Costello.
\newblock {Mitochondrial respiration in hummingbird flight muscles}.
\newblock {\em Proc. Natl. Acad. Sci. USA}, 88(11):4870, 1991.

\bibitem{hoppeler1984scaling}
H.~Hoppeler, S.L. Lindstedt, H.~Claassen, C.R. Taylor, O.~Mathieu, and E.R.
  Weibel.
\newblock {Scaling mitochondrial volume in heart to body mass}.
\newblock {\em Respir. Physio.}, 55(2):131--137, 1984.

\bibitem{robin1988mitochondrial}
E.D. Robin and R.~Wong.
\newblock {Mitochondrial DNA molecules and virtual number of mitochondria per
  cell in mammalian cells}.
\newblock {\em J. Cell. Physio.}, 136(3):507--513, 1988.

\bibitem{wang1997turnover}
R.H. Wang, L.~Tao, M.W. Trumbore, and S.L. Berger.
\newblock Turnover of the acyl phosphates of human and murine prothymosin
  $\alpha$ in vivo.
\newblock {\em J. Biol. Chem.}, 272(42):26405, 1997.

\bibitem{kumei1989reduction}
Y.~Kumei, T.~Nakajima, A.~Sato, N.~Kamata, and S.~Enomoto.
\newblock {Reduction of G1 phase duration and enhancement of c-myc gene
  expression in HeLa cells at hypergravity.}
\newblock {\em J. Cell. Sci.}, 93(2):221--226, 1989.

\bibitem{bogenhagen1974number}
D.~Bogenhagen and D.A. Clayton.
\newblock {The number of mitochondrial deoxyribonucleic acid genomes in mouse L
  and human HeLa cells}.
\newblock {\em J. Biol. Chem.}, 249(24):7991, 1974.

\bibitem{gillespie1977exact}
D.T. Gillespie.
\newblock {Exact stochastic simulation of coupled chemical reactions}.
\newblock {\em J. Phys. Chem.}, 81(25):2340--2361, 1977.

\bibitem{enver2009stem}
T.~Enver, M.~Pera, C.~Peterson, and P.W. Andrews.
\newblock {Stem cell states, fates, and the rules of attraction}.
\newblock {\em Cell Stem Cell}, 4(5):387--397, 2009.

\bibitem{macarthur2009systems}
B.D. MacArthur, A.~Ma'ayan, and I.R. Lemischka.
\newblock {Systems biology of stem cell fate and cellular reprogramming}.
\newblock {\em Nat. Rev. Mol. Cell Biol.}, 10(10):672--681, 2009.

\bibitem{muller2008regulatory}
F.J. M{\"u}ller, L.C. Laurent, D.~Kostka, I.~Ulitsky, R.~Williams, C.~Lu, I.H.
  Park, M.S. Rao, R.~Shamir, P.H. Schwartz, et~al.
\newblock {Regulatory networks define phenotypic classes of human stem cell
  lines}.
\newblock {\em Nature}, 455(7211):401--405, 2008.

\bibitem{graf2008heterogeneity}
T.~Graf and M.~Stadtfeld.
\newblock {Heterogeneity of embryonic and adult stem cells}.
\newblock {\em Cell Stem Cell}, 3(5):480--483, 2008.

\bibitem{huang2007bifurcation}
S.~Huang, Y.~P. Guo, G.~May, and T.~Enver.
\newblock {Bifurcation dynamics in lineage-commitment in bipotent progenitor
  cells}.
\newblock {\em Dev. Biol.}, 305(2):695--713, 2007.

\bibitem{chickarmane2009computational}
V.~Chickarmane, T.~Enver, and C.~Peterson.
\newblock {Computational modeling of the hematopoietic erythroid-myeloid switch
  reveals insights into cooperativity, priming, and irreversibility}.
\newblock {\em PLoS Comput. Biol.}, 5(1):e1000268, 2009.

\end{thebibliography}
\end{document}